\documentclass[aps,prx,reprint,onecolumn]{revtex4-2}
\usepackage{graphicx}
\usepackage{nicematrix}

\usepackage{comment}
\usepackage{xcolor}

\definecolor{myred}{rgb}{0.9176, 0.2000, 0.1373}
\definecolor{myblue}{rgb}{0.1647, 0.3373, 0.4549}

\usepackage{amsmath,amssymb}
\usepackage[colorlinks=true,bookmarks=false,citecolor=blue,urlcolor=blue, linkcolor=myblue]{hyperref} 

\usepackage{physics}
\usepackage{bm}
\usepackage{graphicx}
\usepackage{amsthm}
\usepackage{subfig}
\usepackage{dsfont}
\usepackage{esint}
\usepackage{blkarray}
\usepackage{tikz}
\usepackage{float}
\usetikzlibrary{fit}

\usepackage[ruled,vlined]{algorithm2e}

\DeclareRobustCommand{\appautoref}[1]{%
  \hyperref[#1]{Appendix~\ref*{#1}}%
}

\usepackage{bibunits} 
\defaultbibliographystyle{apsrev4-2}
\defaultbibliography{references-2}

\makeatletter

\newcommand{\hideMainTOCEntries}{%
    \let\savedlpart\l@part
    \let\savedlsection\l@section
    \let\savedlsubsection\l@subsection
    \let\savedlsubsubsection\l@subsubsection
    \let\savedlparagraph\l@paragraph
    \let\savedlsubparagraph\l@subparagraph
    \let\savedlappendix\l@appendix
    \let\l@part\@gobbletwo
    \let\l@section\@gobbletwo
    \let\l@subsection\@gobbletwo
    \let\l@subsubsection\@gobbletwo
    \let\l@paragraph\@gobbletwo
    \let\l@subparagraph\@gobbletwo
    \let\l@appendix\@gobbletwo
}

\newcommand{\showSupplementTOCEntries}{%
    \let\l@part\savedlpart
    \let\l@section\savedlsection
    \let\l@subsection\savedlsubsection
    \let\l@subsubsection\savedlsubsubsection
    \let\l@paragraph\savedlparagraph
    \let\l@subparagraph\savedlsubparagraph
    \let\l@appendix\savedlappendix
}

\makeatother

\begin{document}

\begin{bibunit}
\title{Coherent Control of Wave Scattering via \\Minimal-Parameter Tuning of Complex Spectra}

\author{Ali H. Alhulaymi}
\email{ali.alhulaymi@yale.edu}
\affiliation{Department of Applied Physics, Yale University, New Haven, CT 06520, USA}

\author{Nazar Pyvovar}
\affiliation{Department of Applied Physics, Yale University, New Haven, CT 06520, USA}

\author{Philipp del Hougne}
\affiliation{Univ Rennes, CNRS, IETR-UMR 6164, F-35000 Rennes, France}

\author{Owen D. Miller}
\affiliation{Department of Applied Physics, Yale University, New Haven, CT 06520, USA}

\author{A. Douglas Stone}
\email{douglas.stone@yale.edu}
\affiliation{Department of Applied Physics, Yale University, New Haven, CT 06520, USA}

\begin{abstract}
We introduce and validate a theoretical framework for coherent control of multichannel linear scattering to route waves through complex geometries with multiple scattering.  We show that steady-state perfect routing solutions are achievable at any frequency via tuning geometric parameters so that multiple complex eigenfrequencies coincide on the real axis. The relevant complex spectra describe critically constrained scattering processes (CCONs), where a specific number of generically accessible outgoing channels are inaccessible due to destructive interference. Focusing on electromagnetic waves, we demonstrate in simulations routing and demultiplexing with high discrimination of signals in a multiport chaotic cavity with a small number of tunable scatterers and free-space signal routing in a grating coupler with a similar number of tunable elements in its unit cell. The minimal number of tuning parameters required is predicted by codimension arguments and validated in simulations. A special class of perfect reflection processes are found to be enhanced in the presence of time-reversal symmetry. A similar approach can be used to implement other interesting functionalities, such as isolation, power division, mode conversion and filtering.  The method can be applied to other classical waves and also to quantum matter waves.
\end{abstract}

\maketitle
\addtocontents{toc}{\protect\hideMainTOCEntries}
\section{Introduction}
\subsection{Overview}

Control of linear waves (electromagnetic, matter and acoustic) in two- or three-dimensional multiple scattering geometries without high symmetry is a fundamental problem across many branches of physics and is also of great relevance to device engineering in electromagnetics and acoustics~\cite{Rotter2017LightControl, Mosk2012ControllingMedia, Derode1995RobustScattering}. Coherent control of wave scattering via interference in simple geometries susceptible of analytic treatment (layered(Fabry-Perot), two-channel (Mach-Zehnder) paraxial gaussian optics, gratings ...) has been understood since the middle of the last century or earlier. However, there has been remarkably little progress in our understanding of achievable control in generic two and three-dimensional geometries, due to the lack of a clear mathematical framework for constructing systems which exploit multi-path interference for control of scattering in such systems.

When any such control problem is posed in a non-symmetric, higher-dimensional geometry, the prevailing constructive approach has been black-box or large-scale ``topology'' optimization based on adjoint methods, which exploit thousands or millions of design parameters to produce optimal designs~\cite{ScalableInverseDesign, Christiansen:21, Miller2012PhotonicDesign, Su2020NanophotonicConsiderations}, typically targeting a fixed function and offering little physical insight or {\it in situ} reprogrammability. A complementary line of work has instead sought \emph{fundamental limits} to electromagnetic response, deriving largely geometry-independent bounds on achievable performance directly from the analytic structure of Maxwell's equations~\cite{Miller2026FundamentalLimits, Chao2022PhysicalLimits, Gustafsson2020UpperBounds, Miller2007FundamentalLimit}. This approach supplies insight that brute-force optimization lacks but is non-constructive: it delimits what is possible without prescribing how to achieve it (if it is even achievable). The present work brings these strengths together: a framework that is constructive, like inverse design, yet general and predictive, like the theory of fundamental limits. We show that controlling scattering and propagation of waves via interference, as well as implementing certain information-processing functions, requires a relatively small and \emph{predictable} number of design/tuning parameters, which can be implemented and reprogrammed {\it in situ} over a large frequency range.

Our work was stimulated by the theory of Coherent Perfect Absorption (CPA)~\cite{Chong2010CoherentLasers, Wan2011Time-ReversedAbsorption, Baranov2017CoherentLight} and Reflectionless Scattering Modes (RSM)~\cite{Sweeney2020TheoryModes,Stone2021ReflectionlessTheory,Dhia2018TrappedProblem} which proved that solutions exist to complicated two and three-dimensional impedance-matching problems in the complex frequency plane, and that these solutions can be realized in steady-state (real frequency) by tuning a small number of parameters of the system and the input wavefronts. In the case of CPA this is achieved by trapping an adapted wavefront in a lossy structure, so that the incident energy is transduced, whereas for RSM the adapted wavefront suppresses all backscattering via interference into the chosen incident channels.  The current work builds on, but moves beyond the insights gained from complex reflectionless spectra to optimization problems not described by complex spectra, and provides a full framework for control and optimization of linear scattering with a minimal number of control parameters.

Complex spectra in open systems are well-known for the case of purely outgoing boundary conditions, where the eigenfrequencies (or energies) are referred to as resonances or quasi-normal modes, and they correspond to poles of the scattering amplitudes~\cite{Lalanne2018LightResonances, Kristensen2020ModelingModes, Zhang2020QuasinormalTheory}.  The more recent work has shown that every reflectionless scattering boundary condition also leads to a similar but distinct complex spectrum, with eigenfrequencies (termed R-zeros) in one-to-one correspondence with the resonances.  Off the real axis these are not steady-state solutions but can be excited with transient pulses~\cite{Kim2025Complex-frequencyPhysics, Baranov2017CoherentCapturing}; when they are tuned to a real frequency they are steady-state time-harmonic solutions (termed Reflectionless Scattering Modes). Typically only a single variable system parameter is sufficient to generate multiple R-zero crossings of the real axis in a bandwidth containing multiple resonances; such crossings can be calculated straightforwardly by methods similar to those used for finding resonances. Reflectionless modes were first predicted and demonstrated in simulations~\cite{Sweeney2020TheoryModes,Stone2021ReflectionlessTheory} and later observed in experiments at microwave and optical frequencies~\cite{Sol2023ReflectionlessRouters, Faul2025AgileMetasurface, shaibe2025superuniversalstatisticstopologicalorigins, Jiang2024CoherentModes, PhysRevResearch.7.023090}. Typically the targeted scattering processes are suppressed by 40-70 dB in experiments (and more in simulations).

While the work on CPA/RSM has generated significant attention as a theory of impedance matching, it does not, in itself, provide a general framework for understanding complex optimization problems in wave scattering and control, because it was not obvious how the complexity and number of tuning parameters required would increase with additional constraints, e.g. to control the forward-scattering into selected channels.  As we will see below, adding additional constraints beyond zero reflection rules out the existence of complex spectra associated with such overconstrained scattering problems in order to implement interesting control functions such as routing, demultiplexing, mode conversion and isolation/circulation. Previous theoretical work has not focused on this crucial question of realizing overconstrained scattering in complex geometries (except by brute force optimization).  In this work we show that there is a framework based on generalized complex spectra, supplemented by codimension arguments, which is able to predict the number of parameters required to solve such overconstrained problems by tuning---either in the design process or {\it actively in situ}---a relatively small number of system parameters. We then validate these predictions via simulations through an extensive statistical study in both guided wave and free-space platforms in the presence and absence of reciprocity and time-reversal symmetry.

One application of this theory would be to complement conventional microwave-filter synthesis based on coupled resonator networks and coupling-matrix design~\cite{snyder2021emerging,cameron1999general,cameron2003advanced,Benzaouia2022AnalyticalMetasurfaces}. While such approaches are powerful and can incorporate tunable or switchable elements to enable reconfigurability, they typically rely on spatially distinct resonant elements with engineered pairwise couplings, which constrains the accessible design space for highly flexible, multi-function operation.

Our approach typically utilizes broad overlapping resonances that are not designed around a center frequency and is compatible with strong scattering, allowing the functions to be tuned over an enormous bandwidth. Such a system may eventually be designed to operate autonomously and compensate for perturbations or environmental changes, making them robust and suitable for harsh and/or dynamic environments.

This article will focus on classical waves with applications to electromagnetic waves and information processing functions. However the principles here apply equally to quantum matter waves or single photons, as our results are based on universal properties of scattering matrices. As our ability to control quantum degrees of freedom keeps increasing with the development of modern quantum technology it is quite possible that these principles will become useful in design of quantum information processing systems. 

\subsection{Previous Related Work}
Recent experiments in tunable microwave cavities have explored such systems and demonstrated several of the control functions analyzed here without a full theoretical framework~\cite{Imani2020PerfectInclusions,sol2022meta,Faul2025AgileMetasurface, Sol2023ReflectionlessRouters}. As noted, here we develop a theory to predict the minimum number of parameters required,  and present simulations of several model systems that are realistically related to these experiments. The simulations confirm our minimal parameter predictions, while also demonstrating the ability to reprogram such functions \textit{in situ} over a wide frequency range. Our approach uses only analytic properties of the elements of the scattering matrix, without needing to assume any spatial or temporal symmetries.  Hence it is remarkably general and applies to all systems of relevance in electromagnetics: reciprocal and non-reciprocal, as well as lossless or systems with loss or gain. For routing it is natural to focus on low-loss reciprocal cases, but we present extensive results for other cases as well below, which characterize the regimes of applicability of our approach.

The study of reflection and transmission zeros in systems with two scattering channels (or ports) has a very long history and is well developed in the context of filter design~\cite{Dimopoulos2012AnalogSynthesis}.  Our work focuses on systems with three or more ports/channels and multiple scattering, as do the experiments cited above; additionally, our approach has uncovered topological properties of transmission zeros, relevant to filter and lens design. There have been previous theory and simulations on multichannel quasi-1D disordered waveguides and billiard resonators, with emphasis on transmission zeros of an $N \times N$ transmission matrix, with relevant results reviewed below~\cite{Kang2021TransmissionSystems}.     
An important recent work by Guo et al.~\cite{Guo2023SingularMatrices} is complementary to our own. This work considered a generalization of CPA and RSM, termed coherent perfect extinction (CPE), in which a specific input wavefront in a chosen subset of the scattering channels leads to zero output in a chosen subset of the scattering channels. In this work CPE processes were defined only for scattering at real frequencies and zeros were assumed to exist in a sufficiently large space of tuning parameters.  The work demonstrated interesting topological properties associated with encircling these CPE zeros. In addition the authors presented a codimension argument for the size of the parameter space needed to achieve CPE, which partially agrees with our results below. However this work did not study complex spectra or propose a relationship to optimization and inverse design, nor did it analyze any overconstrained problems, which are the focus of the current work. 
We show below that complex spectra play a crucial role in the tunability of such systems, since, unlike zeros in an arbitrary parameter space, zeros in the complex plane are conserved under perturbations and move continuously. In the next section we define the full set of complex spectra relevant to CPE.  In this work we will only show results on routing and demultiplexing, which are CPE processes, but the same approach can be applied to power splitting, maximal contrast in transmission/reflection and full S-matrix control, which are not. We will present results demonstrating these functions in future work.

\begin{figure}[htp]
    \centering
    \includegraphics[width=0.8\textwidth]{Figure_1_v6.png}
    \caption{A schematic for different kinds of CCONs in a 3-port system and the corresponding submatrix of $S$ (in green), which must have zero determinant. (a) A schematic of a simple 1 port reflectionless mode $(\tilde{R}TT)$, (b) a 2 port Reflectionless mode $(\tilde{R}\tilde{R}T)$ and (c) A ``dark" CCON $(ND\tilde{R}$): port 1 is ``normal", and has both an input and reflected wave, port 2 is ``dark" and has neither input or output, and port 3 is reflectionless. (d) A scatter plot of the zeros corresponding to the different CCON boundary conditions in (a)-(c), including CPA $(\tilde{R}\tilde{R}\tilde{R})$, and resonances $(TTT)$, obtained from the chaotic cavity in Fig.~\ref{fig:MS_fig3}. with 10 tunable scatterers and 5 different orientations. The scale bar on the right shows the mean position on the imaginary axis over 400 random orientations of the scatterers, showing that generically all CCONs lie in  strip of the complex plane between the poles and the full zeros of S for lossless reciprocal scattering (see the \hyperref[title:SM]{SM} for more details where the eigenvalue solver developed is based on an extension of the method proposed in~\cite{Shao1995AnProblems}).}
    \label{fig:MS_fig1}
\end{figure}
\section{Critically Constrained Scattering Modes}
\subsection{Definition of CCONs}
The work cited above on R-zeros demonstrated the existence of discrete complex spectra associated with reflectionless scattering in one to one correspondence with the poles of the scattering matrix. It analyzed generic finite scattering systems with short-range interactions (for simplicity we don't consider monopole charge interactions); the scattering can be between free space or translationally invariant guided modes (channels), modeled as extending to infinity. The same generality applies to the current work. Reflectionless scattering is an example of a type of scattering which we henceforth refer to as critically constrained, because it only admits discrete and (generically) complex frequency solutions.

In contrast, for a standard linear scattering process the solutions typically have a continuous real spectrum: one or more of the incoming scattering channels is filled at a real frequency, $\omega$ or energy, $E$, and the continuity conditions on the relevant wave equation determine a unique outgoing scattered wave at that 
$\omega,E$. For a finite-sized but otherwise arbitrary scattering region, scattering can be truncated to a finite $N_c \times N_c$ scattering matrix, $S$, with elements labeled by the pair of mode indices of the incoming and outgoing channels which are coupled within the scatterer. For generic scattering (scatterers with no spatial symmetries) from a finite region all output channels are accessible in scattering, i.e. no matrix element of $S$ exactly vanishes except possibly at discrete frequencies (once the S-matrix has been appropriately truncated); with discrete symmetries similar statements hold in each symmetry-labeled subspace. Discrete complex spectra arise when certain constraints are placed on the filled output channels, an example being zero reflection back into a chosen set of input channels. This R-zero boundary condition implies that the determinant of a diagonal block (or single diagonal element) of $S$ vanishes. When satisfied, the reflectionless boundary condition implies a solution exists with an input vector of amplitudes of the form ${\bf \alpha} = (a_1 \ldots a_{N_{\rm in}},0,0 \ldots, 0)$ and an output (scattered) vector of amplitudes ${\bf \beta} = (0,0 \ldots ,0,b_{N_{in} +1} \ldots, b_{N_{c}})$. Hence, there are exactly $N_c$ asymptotic scattering channels filled, $N_{\rm in}$ with incoming waves and the complementary $N_c-N_{\rm in}$ with outgoing waves, so that each port is excited in only one direction.  

As a step towards imposing more constraints on the output to route waves, we first  generalize beyond reflectionless scattering to other constrained scattering processes which share the property of having $N_c$ filled channels.
\begin{itemize}
\item 
Definition of a CCON: A critically constrained scattering process or mode (CCON), has exactly $N_c$ zeros (unfilled channels) and $N_c$ non-zero entries (filled channels) in the combined input and output vectors $\alpha,\beta$ of the Scattering Matrix. 
\end{itemize}

We will show that these new processes also have discrete complex spectra and associated eigenfunctions, defining the required input wavefronts. Intuitively, this definition retains the basic mathematical structure as that which led to complex R-zero spectra.  We will justify this statement below.

The set of CCONs as just defined includes all R-zeros and resonances, but also processes in which some scattering channels are completely empty or ``Dark'' (no input or output) and others are ``Normal'' (both an input and a reflected outgoing wave).  Since there are $2N_c$ total input and output channels, our condition is equivalent to having exactly $N_c$ zeros in a $2N_c$-component vector. It follows that for an $N_c \times N_c$ scattering matrix, there exist $(2N_c)!/(N_c!N_c!)$ CCON spectra, $2^{N_c}$ of which are the previously known R-zeros (including resonances and CPA).  Each CCON can be specified by an $N_c$ component vector designating all channels with 1 of 4 letters: $D$ (dark channel), $N$ (normal channel), $T$ (``transmission'' channel, outgoing wave only), and $\tilde{R}$ (reflectionless: input but no output). R-zeros are CCONs which have no $N$ or $D$ channels; all  other CCONs must have an equal number of $N$ and $D$ channels. The number of each of these must satisfy $n_D + n_N + n_T + n_{\tilde{R}}=N_c$. 

Crucially,
\begin{itemize}
\item 
Each CCON frequency is determined by the condition that, at that frequency, a specific square $q \times q$ constraint matrix, $C (\omega)$, projected out of the S-matrix, has determinant equal to zero (where $q = n_D + n_{\tilde{R}}$ is the number of empty channels imposed on the output vector, $\beta$).
\end{itemize}
This property shows that CCONs are a natural extension of the previously known R-zero spectra and will allow one to prove their existence and other relevant properties, as discussed below. CCONs are a subset of all Coherent Perfect Extinction processes as define in Ref. \cite{Guo2023SingularMatrices}, which however did not discuss complex spectra.  CCONs are the only CPE processes that have complex spectra, which is why they can serve as building blocks for more constrained scattering processes.

Schematics illustrating several three-channel CCONs are shown in Fig. (1a-c), labelled as $(\tilde{R}TT)$, $(\tilde{R}\tilde{R}T)$ and $(ND\tilde{R})$ respectively, along with the sub-matrices whose determinants must vanish. Fig. 1d gives an example of calculated spectra for the five distinct types of 3-channel CCONs (neglecting permutation of ports). 

In analogy to R-zero spectra we expect all CCON boundary conditions to generate discrete complex frequency solutions in any real frequency interval in which the system has resonances, independent of the symmetry, dimension or spatial complexity of the system. Similar to resonances the density of these zeros increases with the volume of the scattering region. To prove the existence of all CCON spectra, under the conditions of generic scattering, we note that the all the elements of the scattering matrix are meromorphic functions of frequency in the complex plane, with isolated poles at the resonance frequencies. It follows that  the determinant of the finite-dimensional constraint matrix, $C(\omega)$, the zeros of which are the CCON frequencies, is also a meromorphic function.  Such a function can only vanish at discrete points in the complex plane, or everywhere. Under the condition of generic scattering (all matrix elements non-vanishing except at discrete frequencies), the special case of CCON zeros everywhere is ruled out. Alternatively, for generic scattering one can show that every CCON is determined by a constraint of codimension two (see section IIIA.2 for definitions) in the complex plane, meaning that their spectrum is discrete. 

Using a coupled mode representation of the S-matrix, one can show analytically (see~\appautoref{Existence of CCONs proof} and section I of the \hyperref[title:SM]{SM}) that under reasonable assumptions all CCON zeros (and not just R-zeros) are in one-to-one correspondence with the resonances of the systems, hence with similar density when projected along the real axis. We have developed numerical techniques to calculate all the CCON spectra and have confirmed these properties, including their average density, in full wave systems beyond coupled mode theory; examples of these spectra are shown in the model calculations of Fig. 1d.  Details on the TCMT construction and density of CCONs are given in~\appautoref{Existence of CCONs proof} and sections I and II of the \hyperref[title:SM]{SM}.

The existence of CCON spectra in the complex plane does not require strong mixing of channels in the scattering region. 
However, achieving routing and other functions in scattering will require tuning CCON eigenfrequencies to the real axis, similar to the tuning required for CPA and RSM. The ability to tune CCONs to the real axis does require that the interchannel scattering be sufficiently strong, so that the required interference is realized in steady-state. The low loss case, which is of most interest for applications, is also particularly favorable for effective tuning, due to a critical coupling argument given in the next section.

\subsection{Generalized Critical Coupling: Proximity of CCON Spectra to Real Axis}

Below we will show that routing of waves in multiple scattering can be achieved by manipulating CCON spectra, and in particular by tuning eigenfrequencies to the real axis.  Here we present arguments indicating that CCON eigenfrequencies in the lossless case are clustered near the real axis enabling efficient and reliable tuning to the real axis using standard optimization approaches.  Tuning a single eigenfrequency to the real axis using geometric parameters (instead of loss or gain) has been previously studied for the case of R-zeros, building on the long-established concept of critical coupling (which does not apply rigorously to R-zeros in general, but does apply in a qualitative sense~\cite{Sweeney2020TheoryModes}).
In particular it has been found that the imaginary part of the eigenfrequency of an R-zero is determined in a statistical sense by the coupling balance between the chosen input and output ports.  In TCMT, in the single resonance limit, there is an explicit scalar formula for this, which constitutes a multichannel generalization of the ``critical coupling" condition~\cite{Sweeney2020TheoryModes}. Specifically, filled output channels appear as scattering loss, and filled input channels as an effective gain; in the more exact S-matrix theory of R-zeros this property is still maintained on average~\cite{Sweeney2020TheoryModes}.
Therefore, if the input channels are more numerous or more strongly coupled than the output channels, then the imaginary part of an R-zero will be positive (above the real axis), consistent with the fact that full S-matrix zeros must be above the real axis in the absence of absorption. In the opposite case of weak input coupling, the R-zeros will be below the real axis on average, and in both cases there will be significant statistical fluctuations~\cite{Sol2023ReflectionlessRouters}. When input and output couplings are roughly in balance, the distribution of R-zeros is centered on the real axis, and an R-zero can be tuned to lie on the real axis at any frequency relatively easily. 
This has been validated in the relatively low-Q open cavities we consider here, where transmission is through multiple overlapping resonances. 

For the CCONs with dark and normal channels it is not immediately obvious how to apply this qualitative reasoning.  As seen in Fig. 1d, on average, for a system with negligible absorption, the poles will be the eigenfrequencies farthest away from the real axis in the lower half plane and the (complex conjugate) zeros of the full S-matrix will be farthest away in the upper half plane.  As introduced above, CCONs can have singly-filled ports ($\tilde{R}$ or $T$), doubly filled ``normal" ports ($N$) and dark ports which are unfilled ($D$). Following the reasoning about R-zeros, an $\tilde{R}$ port contributes an effective gain and a $T$ port an effective loss moving the frequencies respectively up or down in the complex plane. A reasonable conjecture is that, on average, a dark port contributes neither effective gain nor loss (since it has neither input or output), and that a normal port, having both, will also have no average influence on the imaginary part of the CCON frequency. Thus, in estimating the imaginary part of the CCON frequency, one can omit all $ND$ pairs and just look at the difference, $n_T- n_{\tilde{R}}$ of the number of remaining $T,\tilde{R}$ ports to determine their average imaginary part. We assume here that all ports are equally coupled in a statistical sense; when this isn't the case one includes the strengths of coupling in the obvious manner.

Hence, the average imaginary part of the reflectionless four-port CCONs $(\tilde{R}\tilde{R}TT)$ and $(NDND)$, which are balanced ($n_T- n_{\tilde{R}}=0)$, should both be zero. This qualitative argument explains why all the CCON spectra tend to lie in the region of the complex plane between the poles and zeros, as is also the case for R-zeros.  For a general scattering matrix based on a random matrix model, we can validate this intuitive argument by analytically calculating the average value of the imaginary part of the CCON eigenfrequencies:
\begin{equation}
    \langle \text{Im} \,\omega_{\rm CCON}\rangle = \frac{\Gamma}{N_{c}}(n_{\tilde{R}}-n_T)
\end{equation}
Here $\Gamma$ is half the average resonance width defined from the coupling matrix $W$ by $\mathrm {Var}(W_{\alpha \beta}) = 2\Gamma/N_{c}$. The expression above is written for the mean over the entire spectrum, and is valid for weak coupling. We expect a similar expression 
 to hold for the local mean as well (see section II of the \hyperref[title:SM]{Supplementary Material} for more details). Full detail on the RMT model and numerical tests thereof are given in section II of the \hyperref[title:SM]{Supplementary Material}.
 
In the next section we study tuning of eigenfrequencies to the real axis to realize control functions. As noted, this is expedited by the fact that they are generically confined in a strip of the complex plane centered on the real axis for the low-loss case. It was previously studied~\cite{Sol2023ReflectionlessRouters} and understood that when the system is greatly overdamped through strong absorption the median of the distributions of R-zeros (now generalized to all CCONs) can fall well below the real axis making tuning to the real axis more difficult.
Nonetheless such tuning is possible in a highly parameterized cavity and demonstrated both in simulations and experiments~\cite{Sol2023ReflectionlessRouters}. This will be discussed further below in the context of codimension estimates for the minimal parameter set to achieve routing.

\subsection{Routing as a Coincidence of CCONs}

 Routing of waves can be defined as zero reflection of the waves back into the chosen input ports and zero transmission from each input port to a subset of the other ports (and falls under the definition of CPE). Since routing adds constraints beyond zero reflection, the problem is overconstrained and routing solutions are not CCONs; hence they don't have an associated spectrum in the complex plane
 (excluding the two-port case, where zero reflection routs waves trivially into the other port).
 We will focus on systems with $N_c \geq 3$ and, initially, on systems with
 negligible absorption. For the three-port case, the only routing solution is $(\tilde{R}TD)$ (and its permutations), corresponding to input in port one with zero reflection ($S_{11}=0$), and output solely in port two ($S_{31}=0$).  Consistent with the above discussion, this CPE process has a $D$ channel not paired with an N channel and isn't a CCON; it is overconstrained, and doesn't have a complex spectrum. As noted, such overconstrained CPE processes have not been studied in previous theoretical work. However, there do exist CCONs satisfying either the first or second constraint: in our notation they are $(NTD)$, which satisfies $S_{31}=0$, and $(\tilde{R}TT)$, which satisfies $S_{11}=0$.  Both have a complex frequency spectrum where the relevant constraint is obeyed. If, by tuning the geometric parameters of the system, we can cause two of these distinct CCON frequencies to coincide at a chosen frequency on the real axis, by definition we will have realized the routing solution we seek. This reasoning generalizes, and any overconstrained CPE process can be realized by creating a {\it coincidence} of a number of CCONs eigenfrequencies on the real axis. Note that this routing function requires a coincidence of both an R-zero and a dark CCON, which is a generic feature. Generically, the number of eigenfrequencies that must coincide to realize a given CPE process is $n_{co} = n_D-n_N + 1$ (the number of system parameters required to generate such coincidences is analyzed below).

\begin{figure}[htp]
    \centering
    \includegraphics[width=0.8\textwidth]{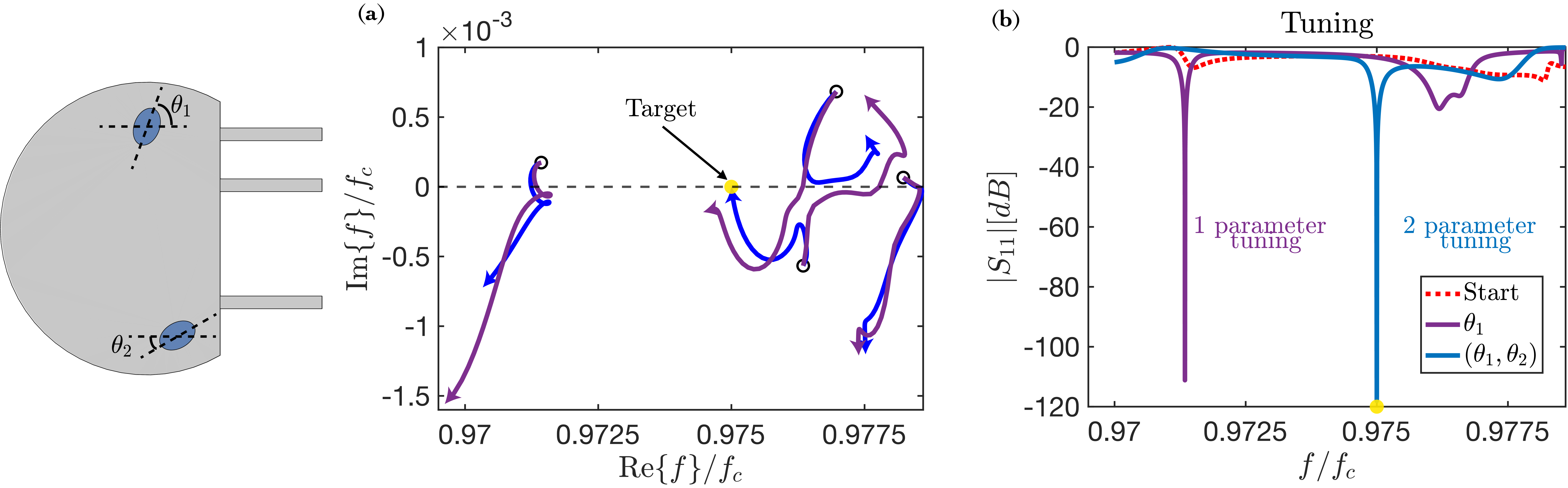}
    \caption{R-zero tracking in a tunable chaotic cavity (inset) with two ellipsoidal scatterers of variable orientation, demonstrating tuning of an R-zero to a target real frequency (yellow dot) for an arbitrary initial configuration. (a) Tuning a single ellipsoid orientation (purple curves) will generate real frequency reflection dips, but not at the target frequency; optimizing with both scatterers (blue curve) gives a high quality solution.  (b) Reflection coefficient of the cavity for three cases: (Solid blue curve) configuration optimized to show reflection dip at the target frequency, $0.975f_c$, where $f_c$ is the cutoff frequency of the waveguides. (Solid Purple curve) reflection spectrum corresponding to single parameter tuning at the configuration when an R-zero passes through the real axis randomly. (Red dashed line) Reflection spectrum at the starting point, showing no deep reflection dips.}
\end{figure}

\subsection{Topological protection of CCONs}

Our strategy for realizing routing and other useful scattering processes assumes that the discrete CCON eigenfrequencies are conserved and flow continuously in the complex plane as parameters of the cavity are varied. Since the CCON frequencies correspond to zeros of the determinant of a constraint matrix, they are zeros of a complex meromorphic function of frequency. 
Any zero of a meromorphic function has a local topological winding number associated with encircling that zero (and no other zero) in an appropriate parameter space.  However, for a general function, such a zero can annihilate with another zero of opposite winding number, which will still conserve the global winding number of the system, but changes the number of zeros available for functionalization.  We are focused here on tuning zeros in a special parameter space, the complex frequency plane, via controlled perturbations. Since all the matrix elements of S are analytic functions of $\omega$ except at the discrete set of poles of S, the determinant of any sub-matrix of S is also analytic in $\omega$, except on the same discrete set, generically with a non-vanishing linear expansion around the zero (i.e., not at an exceptional point). It is known in general that the zeros of analytic functions have winding number $+1$ and we review this for CCON zeros using the Cauchy--Riemann relations in section VIII of the \hyperref[title:SM]{Supplementary Material} (see also~\cite{Roe2015WindingAnalysis, Milnor1965TopologyViewpoint}); hence they cannot annihilate pairwise.  Therefore, CCON eigenfrequencies are conserved in the complex plane under parametric variation (except for the special case of annihilation with a pole of the S-matrix, which is negligible in this context).  This differentiates the complex plane as a special parameter space, different, e.g., from the space of geometric parameters (see section VIII in the \hyperref[title:SM]{SM}), where only global winding numbers are conserved, but zeros are not. Another qualitative difference to be discussed below is that certain CCONs are not discrete in the generic parameter space, and instead lie on continuous curves.
Excluding this case, we find that when we fix the frequency and vary geometric parameters, generically we observe regular
creation and annihilation of zeros in pairs of opposite helicity in such a generic parametric subspace, see section VIII in the \hyperref[title:SM]{SM}. This annihilation process is very similar to the annihilation of Bound States in the Continuum (BICs) of opposite helicity found in photonic crystal lattices as lattice parameters are varied~\cite{Hsu2013ObservationContinuum, Zhen2014TopologicalContinuum, Hsu2016BoundContinuum}.
While the topological protection of S-matrix zeros has long been noted, the fact that R-zeros (and their generalization to CCONs) are absolutely conserved and cannot vanish via pairwise annihilation has not been emphasized in prior work, to our knowledge. This property underlies our ability to succeed in functionalization and coherent control. 

\section{Scattering Control via CCONs}

\subsection{Minimal Required Parameter Set}
\subsubsection{R-zero and CCON tuning}

We now will discuss the minimal parameter set needed to achieve routing.  We start with tuning a single CCON to the real axis: in the R-zero theory it has been shown that in general tuning an eigenvalue to reach the real axis requires just a single tuning parameter (e.g., adding uniform loss for CPA, or tuning a mirror reflectivity for R-zeros). In fact, since there are many R-zeros with the same density (projected onto the real axis) as the resonances, tuning a single parameter can cause many axis crossings, as is shown in Fig. 2.  However such single parameter variation does not allow us to target a specific frequency. A more useful objective function is to realize a reflectionless state (or any CCON) {\em at a specified frequency}, $\omega_0$, on the real axis. Since here we are requiring that the complex determinant of the constraint matrix vanish at a specific frequency, intuitively two parameters at least are needed. Indeed this is found to be sufficient for the example shown in Fig. 2, where we show the flow of the R-zeros of a chaotic cavity with two tunable scatterers (see schematic). The sufficiency of two-parameter tuning for targeted placement on the real axis for all CCONs is supported by a thorough statistical study shown in Fig. 5 below, where the details of the models used are also given. While in principle possible, we do not find it efficient to optimize by tracking CCONs in the complex plane, hence we use gradient-based optimization methods which work well (see section IX in the \hyperref[title:SM]{SM}).  In the next section we provide codimension arguments for the minimal parameter set needed. This approach is able to accurately evaluate the effect of symmetries on the minimal parameter set. It predicts that for the lossless reciprocal case most CCONs require two tuning parameters to realize as steady-state solutions at a specified frequency, as found in Fig. 2 above. But, remarkably, there does exist one type of CCON that requires only a single efficient tuning parameter to achieve the same functionalization, as confirmed by the data of Fig. 5. 

\subsubsection{Codimension Argument: Generic Case}
A systematic approach to minimal parameter estimation is based on the conjecture that the minimum number of effective tuning parameters required to achieve a certain function is equal to the codimension of the matrix space in which that particular set of constraints is satisfied~\cite{Guo2023SingularMatrices}. This conjecture is tested and verified in simulations over three different symmetry classes of S-matrices (see Fig. 5 below), but can be violated (see below).  Similar codimension arguments were given previously in~\cite{Guo2023SingularMatrices}, but without any statistical study of the conjecture relating codimension to number of tuning parameters, and without taking the effects of symmetry into account, which change the codimension of certain CPE processes in the lossless reciprocal case.

 The space of all possible S-matrices satisfying the global symmetries of interest (e.g. unitarity or reciprocity) form a smooth real manifold of finite dimension, equal to the number of independent real parameters necessary to specify each member of the matrix ensemble. The codimension associated with a CPE process in that space is the difference between the dimension of the full space and the sub-manifold on which the CPE constraints are satisfied.  
 
We consider four ensembles of S-matrices: lossless, reciprocal S-matrices (unitary symmetric), lossless non-reciprocal S-matrices(unitary, non-symmetric), and both cases without the unitarity constraint (i.e. with loss or gain).  In calculating the codimension in the presence of symmetries, we must assume that $\omega$ is real, as unitary S-matrices on the real axis will not be unitary for complex frequencies.

For all except the unitary symmetric case, straightforward calculation of the codimension associated with all CPE processes show that it is two for all CCONs (see section IV in the \hyperref[title:SM]{SM}), and is increased by two for every additional constraint (dark channel added) in an overconstrained CPE process.  Based on the conjecture stated above, this implies that the minimal number of efficient tuning parameters, $n_p$, needed to reach CPE for these three ensembles is: 

\begin{equation}
   n_p = 2(n_D-n_N + 1),
\end{equation}
valid for all CPE processes in all but the unitary symmetric ensemble. We will show below that in the unitary symmetric (lossless reciprocal) case, the same formula holds for the subset of CPE processes corresponding to routing, since for efficient routing we always want reflectionless inputs. Thus, for routing, the relevant codimension is the same in all ensembles, and if our conjecture is correct, then the minimal number of parameters required is always given by Eq. (2).  Of particular interest is the full routing process, reflectionless input and unit output in a single of the complementary ports, $(\tilde{R}D \ldots DT)$, for which $2(n_D - n_N + 1)= 2(N_c-1)$, predicting that four-port routing requires six parameters (see confirmation below, Fig. 5).

 The codimension prediction in Eq. (2) agrees with that found previously by Guo\cite{Guo2023SingularMatrices}, except, as noted, in the unitary symmetric case, however it has never been verified previously for overconstrained solutions.

 \subsubsection{Codimension Argument: Lossless Reciprocal Case}
 For the unitary symmetric case, however, not all CPE processes have the same codimension. There exist interesting processes which do not obey Eq. (2), specifically CPE processes which involve only $n_D$ dark and $n_N$ normal channels, with $n_D \geq n_N$ (and no singly occupied channels). Such processes can be described as perfect reflection into a subset of channels, or equivalently, as zeros of a certain transmission matrix.  Transmission zeros are important quantities in filter and metalens design. We find (see section IV.B in the \hyperref[title:SM]{SM}) that for such processes the combined symmetries reduce the codimension by exactly a factor of two compared to (2):
 \begin{equation}
   n_p^{PR} = (n_D-n_N + 1),
\end{equation}
making such processes achievable with fewer tuning parameters (the superscript denotes perfect reflection). 
Note that only for perfect reflection is the time-reverse of such a process also a perfect reflection process; this is why such processes are physically different from other CPE processes.

The fact that such processes are easier to optimize due to time-reversal symmetry is a new and important result of our work.
It is closely related to a previously known result from Kang et al.\cite{Kang2021TransmissionSystems}, who  showed that the complex spectra for perfect reflection into half the channels ($n_D=N_c/2$ and $n_N=N_c/2$), which is a CCON, must occur in complex conjugate pairs or on the real axis (a simple proof of this is given in section IV of the \hyperref[title:SM]{SM}, using only time-reversal symmetry). However ref.\cite{Kang2021TransmissionSystems} did not consider the implications of this finding for parametric tuning on the real axis. According to our formula (3), this process has {\it codimension one}, and should be achievable at a given frequency by tuning a single parameter, unlike all other CCONs, which have codimension two, and require two tuning parameters. We will see below that these predictions are confirmed by simulations for the lossless reciprocal case. Note that while transmission zeros have codimension one in a generic parameter space, they remain codimension two in the complex plane, consistent with discreteness of CCONs, precisely because the S-matrix is not unitary off the real axis. 

This is a surprising and striking result: one can ``turn off", via destructive interference, all $N_c/2$ output ports, by tuning a single parameter, without the help of any spatial symmetries. Alternatively, by simply scanning frequency in such a system, arbitrarily deep transmission zeros will be found with no system tuning at all. And, in case of low loss systems, many transmission zeros will be found just below the real axis. This is consistent with the experimental findings of refs~\cite{Faul2025AgileMetasurface,Sol2023ReflectionlessRouters} that transmission zeros are the easiest CCONs to tune to the real axis. 

An intuitive explanation for this is as follows: assume a reciprocal lossless system and that isolated real eigenfrequencies exist at some $\omega$ for some values of two tuning parameters $\theta_1,\theta_2$. Changing the value of $\theta_1$ can only move the frequency along the real axis (due to the isolation of the solution) and the same is true of tuning $\theta_2$; hence when we tune $\theta_1$ and move the zero away from $\omega$, we can then tune $\theta_2$ by just the right amount, so as to move it back to $\omega$. It follows that this transmission zero at $\omega$ can exist along a curve in this two dimensional parameter space $\theta_1 -\theta_2$; hence these zeros are not isolated in that space, and the codimension of S-matrices satisfying the constraint $det(T)=0$ is one. An example of such curves for our model is shown in Fig. S2 of the \hyperref[title:SM]{SM}. 

  Our codimension analysis goes further than just describing the balanced perfect reflection CCON, with codimension one, to describing all overconstrained perfect reflection processes. Such unbalanced perfect reflection, with fewer filled input channels than potential output channels, are not CCONs, and are harder to realize than the balanced process. But they are still easier to achieve by a factor of two than other CPE processes with the same number of constraints. A striking special case of this is perfect reflection into a single input channel, $(ND \ldots D)$, which requires only $N_c-1=n_D$ parameters to achieve as compared to perfect routing, i.e. transmission into a single output channel, ($\tilde{R}D \ldots DT$) which requires $2(N_c-1)$. This prediction is confirmed in Figs. 5,6. This implies that there is a positive correlation of the motion of CCONs, so that it is easier to create coincidences on the real axis then if they were uncorrelated and each required two additional parameters.

 The codimension calculations, while rigorous, do not confirm the conjectures given in Eqs. (2),(3) as to the minimum number of physical tuning parameters needed, i.e. they don't guarantee that such solutions can always be reached by tuning the minimal number of physical parameters in a specific geometry, and it is clear that  independent and efficient tuning parameters are required. 
 However these arguments imply that if one doesn't have at least the specified number of tuning parameters for a particular control function, finding good solutions is highly unlikely.  A major result of our work is to demonstrate that there are physically realistic, tunable, guided wave and free space model systems 
that realize all CPE processes while using the minimum size parameter set predicted by codimension arguments. 
The results of these statistical studies are presented in Figs. 5,6; the guided wave models are closely related to the experiments in Refs.~\cite{Sol2023ReflectionlessRouters, Faul2025AgileMetasurface}. For concreteness, we first present examples of routing and demultiplexing solutions found by optimization within one of these models, guided by the formula for $n_p$ given above.

\begin{figure}[htp]\label{Routing fig}
    \centering
    \includegraphics[width=0.8\textwidth]{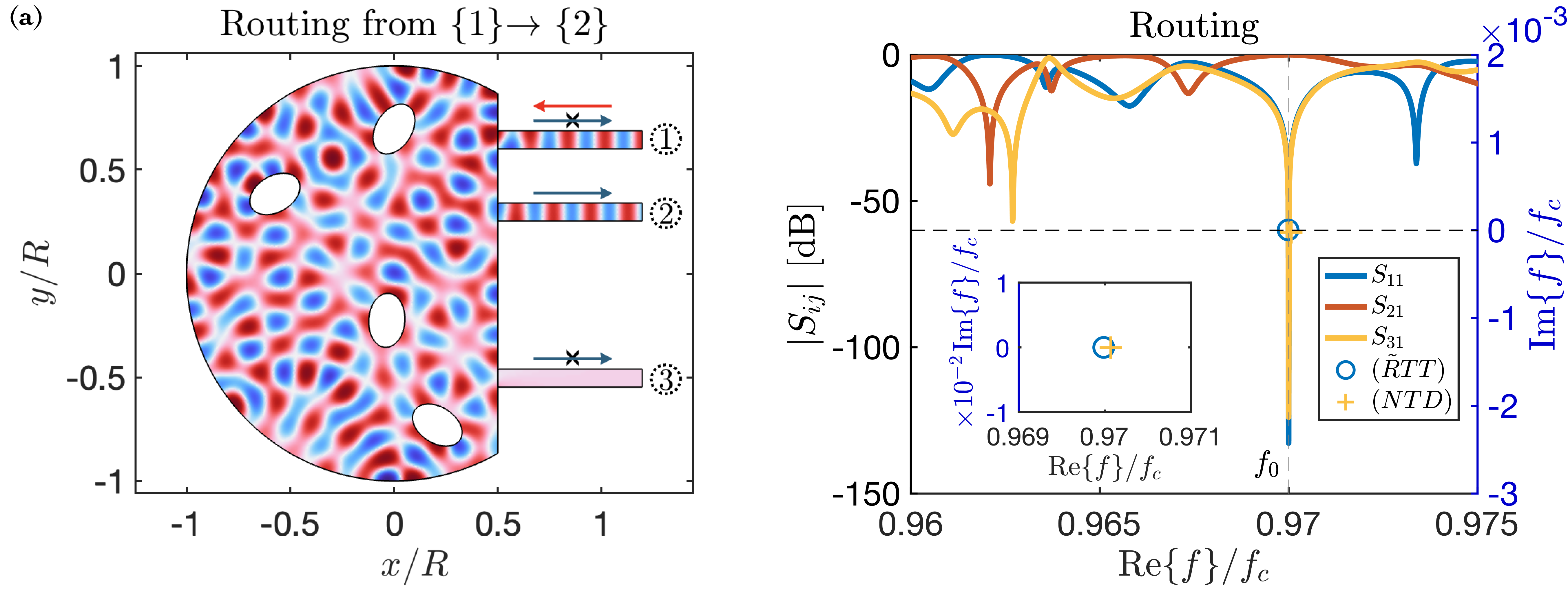}
    \caption{Example of a routing solution (see arrows) found by four parameter optimization, the minimum number of required parameters predicted by the theory, see Eq. (2). (a) Heat map of the optimized EM field, $H_z$, showing routing from port 1 to port 2 with discrimination $ > 75 \; dB$. (b) Scattering matrix elements when exciting channel 1. Inset: Blow up of a region in the complex frequency plane near the real target frequency, showing the predicted coincidence of CCONs, $(\tilde{R}TT)$ and $(NTD)$, to accuracy$ <10^{-7}$.  }
    \label{fig:MS_fig3}
\end{figure}

\subsection{Examples: 3-port Routing and Demultiplexing}

We model a ``D-shaped" microwave cavity with tunable scattering elements embedded within it via a 2D structure with a boundary given by a truncated circle (proven to generate fully chaotic ray dynamics upon specular reflection even in the absence of the embedded scatterers~\cite{Ree1999ClassicalCut,Doya2001OptimizedAmplifier}). As shown in section IV.C of the \hyperref[title:SM]{SM}, simulations with rectangular cavities and similar scatterers perform just as well, suggesting that a few scattering elements may be sufficient without choosing a ray-chaotic cavity shape, although this point still requires further study. In the simulations of Figs 5(a) the walls of the cavity are perfectly conducting except where a number of single-mode waveguides are attached on the flat part of the boundary; in Fig. 5(b) a finite conductivity in the wall is introduced to study the effect of loss. Inside the cavity there are a variable number of wavelength scale perfectly conducting ellipsoids (the scatterers) that are free to rotate, hence altering the scattering of waves injected from the input ports. The model mimics the experimental system of Ref.~\cite{Faul2025AgileMetasurface}, where screws are inserted into a cavity of the same boundary shape at variable depths to modify the scattering behavior. In the experimental work the configuration of screw depths is used for optimization, and that work has found that good, narrowband routing solutions are obtainable experimentally. However those experiments used significantly more parameters than our theory suggests are required, and didn't study more highly constrained functions, such as demultiplexing.
 
 In our model the rotation angles of the ellipsoids are the optimization parameters.  Our prediction is that we will need at least four tunable scatterers to move the appropriate pair of CCONs to the target frequency and achieve high quality 3-port routing. In Fig. 3 we show a typical optimization result, achieving $\sim 108\; {\rm dB}$ suppression of the reflection and undesired transmission with four-parameter tuning. The scattering matrix calculations here were performed with a COMSOL-based code (the statistical data of Fig. 5 used a boundary integral method based custom code~\cite{chunkie}, which has higher accuracy, and gives quantitative agreement). This routing solution was obtained using a relatively simple two-stage optimization procedure: an unconstrained quasi-Newton search implemented with MATLAB's \texttt{fminunc} routine, followed by refinement with Nelder-Mead simplex algorithm implemented with \texttt{fminsearch} routine in MATLAB~\cite{MATLAB_R2024a, OptimizationToolbox_R2024a} (see section IX in the \hyperref[title:SM]{SM}). The data in Fig. 3 are typical; we generated many solutions based on different initial spatial positions and orientations of the scatterers with similar results. Importantly, our similar study with only three and two tunable scatterers found only $\sim 14\;{\rm dB}$ and $\sim 32\; {\rm dB}$ of suppression, respectively. 
 
\begin{figure}[htp]
    \centering
    \includegraphics[width=0.8\textwidth]{Figure_4_v6.png}
    \caption{Demonstration of reprogrammable demultiplexing in chaotic cavity with 8 tunable scatterers (as predicted by theory). The target frequencies are $f_1 = 0.965 f_c$, at $f_2 = 0.975 f_c$. The data shown in fig. (a) correspond to the top routing schematic, with $f_1$ signal routed to port two and $f_2$ signal routed to port 3. The scattering matrix elements (left axis) show deep dips for $S_{11}$ at both $f_1$ and $f_2$, and for $S_{31}$ at $f_1$ and for $S_{21}$ at $f_2$. Symbols (right axis) are the CCON spectra (see legends) with a coincidence of $(\tilde{R}TT)$ and $(NTD$ at $f_1$ and of $(\tilde{R}TT)$ and $(NDT)$ at $f_2$, as predicted by theory. Panels (c-d) show heat maps of the EM field (Re\{$H_z$\}) with the desired routing patterns at $f_1,f_2$ for case (a). 
    (b) Shows similar data to (a) for the bottom routing schematic, where the roles of ports 2 and 3 interchanged (we don't show field maps for this case). }
\end{figure}

 In Fig. 4 we validate our framework for the more demanding functionality of demultiplexing, which should be achievable in the same three-port cavity if the number of scatterers is increased to eight. This is just routing at two distinct frequencies: in terms of CCONs it is $(\tilde{R}TT)$ and $(NTD)$ at $f_1$ and $(\tilde{R}TT)$ and $(NDT)$ at $f_2$, requiring double the four scatterers required for routing at a single frequency.  The system is linear, so we are able to simultaneously send in signals at $f_1$ and at $f_2$ through port one and the two signals will be distributed into ports $2$ and $3$ respectively, with high discrimination and no reflection to perform demultiplexing.  As expected, we are able to find solutions (Fig. 4), with $\sim 40\;{\rm dB}$ suppression of unwanted scattering at two different frequencies.  Demultiplexing with similar discrimination has already been experimentally demonstrated in a microwave cavity with tunable scattering via a metasurface with hundreds of binary tuning parameters~\cite{Sol2023ReflectionlessRouters}, but with a $\sim 20\;{\rm dB}$ overall loss due to absorption. However, Low-loss demultiplexing as we find in our simulations, which would be critical for applications, has not yet been demonstrated in any experimental system similar to the one modeled here. Neither routing or demultiplexing has been demonstrated with as few tuning parameters as we use here.

 We emphasize that the frequencies chosen to demonstrate routing and demultiplexing were chosen arbitrarily. Not only can we choose any frequencies for the desired function over a large bandwidth, we can also interchange spatial pairing, e.g. have demultiplexing from ports $1 \to 3$ at $f_1$ and $1 \to 2$ at $f_2$, for a different configuration of the same scatterers, as we show in Fig. 4b.  Thus, as emphasized in~\cite{Faul2025AgileMetasurface, Sol2023ReflectionlessRouters}, a system such as this has broadband {\it in situ} reprogrammability.
 In Figs. 5 of the next section we show a more extensive statistical study of the model for the richer example of four-port routing.

\subsection{Statistical tests of minimal parameter set}

\subsubsection{Statistical tests: tunable chaotic cavity}

 The examples above of coherent control of scattering achieved with the minimal number of parameters are encouraging. However to demonstrate the validity, robustness and universality of our predictive framework, we performed a statistical study of achievable performance versus number of tuning parameters over many different CPE processes for both guided wave models and for a free-space grating coupler described in the next section. We find excellent agreement with our minimum parameter predictions for up to six tuning parameters (guided wave examples) and ten tuning parameters (grating examples).
 
 First we study guided wave, four-port chaotic cavity and network models, realized in 
 three different symmetry ensembles relevant to experiment: lossless reciprocal systems, lossless non-reciprocal systems and lossy reciprocal systems.   For the lossless reciprocal case and lossy reciprocal cases we use the D-cavity model discussed above, with orientation angles of the ellipsoidal scatterers as the tuning parameters and with the spatial position of the scatterers varied to generate a statistical ensemble of S-matrices.  For the non-reciprocal case we use the ``quantum graph model''~\cite{Kottos1997QuantumGraphs, Kottos2003QuantumScattering}, which is easily realized in experiments in microwave networks. In the graph model the tunable parameters are scattering phase shifts along the ``bonds" of the network, and the bond lengths of the networks are varied to generate the statistical ensemble.  For each CPE process we defined appropriate figures of merit (FOM) related to the suppression of the unwanted transmission and/or reflection processes (see section IX in the \hyperref[title:SM]{SM} for details) and averaged them over the results of multiple optimizations with results shown in Fig. 5 (where we also indicate their standard deviations). We study up to six-parameter control functions, the most demanding being four-port routing. 
 In addition, in the \hyperref[title:SM]{Supplementary Material} we use the quantum graph model~\cite{Kottos1997QuantumGraphs, Kottos2003QuantumScattering} to check our predictions for up to ten tuning parameters for the lossless reciprocal case. 
 With up to eight tuning parameters in a four-port cavity, we can study eight different CPE processes (see legend in Fig 5). The different CPE processes in the lossless reciprocal case, Fig. 5a. correspond to two-port perfect reflection, reflection/transmission zeros, one-port perfect reflection, partial routing (coincidence of two CCONs) and full routing (coincidence of three CCONs), requiring, 1,2,3,4 and 6 parameters respectively. Note particularly that the two symmetry-enhanced perfect reflection processes obey Eq. (3) and not Eq. (2). All results show excellent suppression of the FOM, with a minimal number of parameters in agreement with theory.

  \begin{figure}[htp]
    \centering
    \includegraphics[width=0.9\textwidth]{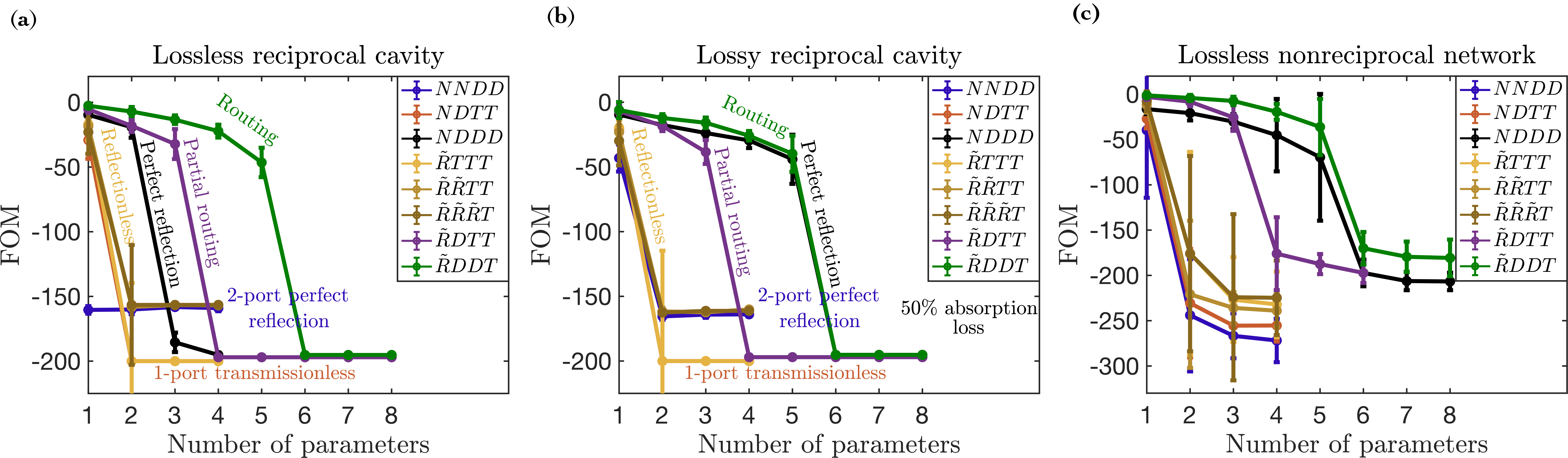}
    \caption{Statistical study of the variation of appropriate Figures of Merit (FOMs) for different CPE processes computed for a four-port chaotic cavity (two-port perfect reflection, reflection zero, single transmission zero, one-port perfect reflection, partial routing, full routing, plus other CPE processes enabled by symmetry breaking in b,c) when optimized with a variable number of tuning parameters. Average performed over ensemble of different chaotic cavities (a,b)) and chaotic networks (c). Data in (a) is for lossless reciprocal chaotic cavity model and in (b) for same cavity with 3 {\rm dB} of total absorption loss due to finite conductivity in the walls. In (c) we studied lossless non-reciprocal network models.  In all cases excellent performance is found for the minimum number of parameters predicted by Eqs (2) and (3). (The data in (a) was obtained from an ensemble of 440 realizations; (b) was obtained using 480 realizations and (c) was obtained from an ensemble of 2100 realizations. Details of models and averaging procedure are given in section IX in the \hyperref[title:SM]{SM}).}
\end{figure}

 For the lossy reciprocal cases (5b) and the lossless non-reciprocal case (5c), the codimension arguments predicts a qualitative change in our results, that all CPE processes will now require an even number of tuning parameters as given by Eq. (2), corresponding to one, two or three coincident CCONs.  In both cases perfect reflection is no longer aided by time-reversal symmetry and both full routing ($\tilde{R}DDT)$ and one-port perfect reflection (NDDD) should now require the {\em same} number of parameters, six, as we find Figs (5b),(5c) (for the lossy case, perfect reflection is a combination of reflected and absorbed flux).  Overall, the agreement for all CPE processes is very good for both cases. To interpret the results one should bear in mind that the theory indicates that there is no lower bound to the FOMs; the closer the eigenvalues come to the real axis and the closer they come to being exactly coincident, the greater the suppression.  Hence the critical feature is at what number of control parameters does the FOM ``dives" down below a threshold (e.g., $-150 {\rm dB}$), and not the ultimate floor. It is also important that we have plotted median FOM and not mean, which could be skewed lower by a few very small values. The results thus show that one should expect to be able to achieve excellent FOMs in all three cases for almost any initial placement of the scatterers. Finally, of course in experiments various noise sources will likely limit the minimum achievable FOM; however in the experiments upon which our cavity model is based they reported typical suppression $> 60 {\rm dB}$, for CCON FOMs and suppression $\geq 40 {\rm dB}$ for three-port routing FOMs, which is quite good, and likely capable of further improvement.
 
The results for the lossy reciprocal case are for the case when the conductivity is chosen to generate $50\%$ $(3\,{\rm dB})$ of total scattering loss.  We find (see section IV.D in the \hyperref[title:SM]{SM}) that when we increase the loss to $10 dB$ it becomes harder to achieve a good FOM for the most overcoupled processes, $(\tilde{R}TTT)$, e.g. one-port reflectionless CCON. This now requires three parameters to achieve instead of only two. 
The qualitative reason for this is understood; above critical coupling all of the CCONs are pulled down well below the real axis (in contrast to Fig. 1d above), making it harder for them to be placed on the real axis via parameter tuning.  Aspects of this physics were studied and confirmed previously, both theoretically and experimentally in our prior work~\cite{Sol2023ReflectionlessRouters}.  We do not think this effect is problematic for applications of our approach to devices, as such high absorption loss would not be acceptable in the first place. Our results for $50\%$ loss make it clear that the minimal parametrization we predict is not adversely affected by reasonable loss. In addition, the results for the lossless non-reciprocal graph model make it clear that our approach can be viable for reprogrammable isolators and circulators.  Here the main practical problem is introducing large non-reciprocity without also generating too much loss. As noted above, realistic simulations presented elsewhere \cite{yuan2026synthesizing}  show that in a similar chaotic cavity with ferrite inclusions isolation $ > 100 \rm{dB}$ should be achievable.

\subsubsection{Free-space grating model}
The geometry of our guided wave models is challenging to employ at higher than microwave frequencies, due to the difficulty of generating strong interchannel scattering without too much leakage loss.  However here we show that our codimension-based minimal parameter prediction will apply and be useful in other very different geometries. We consider a one-dimensional three-layer periodic lossless dielectric grating in free space, coupling into a glass dielectric (to break up-down symmetry). Here we use a step-wise spatially-varying dielectric function as our design parameters.  Each layer is divided into up to four segments with a chosen real dielectric constant in each segment ranging between 1-12, affording us up to twelve tuning parameters (we simply combine segments when using fewer). The thickness of each layer is $\lambda/2$, where $\lambda$ is the free space wavelength, and the period is $\Lambda = 1.2\lambda$. Taking into account that the average index of refraction for the optimized solutions is $\sim 2$ the system has an area equivalent to roughly eight internal squared wavelengths, which we find is sufficient for generating multiple scattering and interchannel interference. The only propagating channels for this system are the $(0,\pm 1)$ diffraction orders in both the reflection and transmission, yielding a six-channel S-matrix.  We define the S-matrix here using the incoming channel ordering $\{-1,0,+1,-1_g,0_g,+1_g\}$ for the columns of $S$ and the outgoing channel ordering $\{+1,0,-1,+1_g,0_g,-1_g\}$ for its rows, where $g$ describes the glass side diffraction orders.  This convention pairs each incoming channel with its time-reversed  outgoing counterpart, and therefore represents reciprocity as a symmetric  S-matrix.  The CPE processes can then be characterized in exactly the same way as  in the chaotic cavity, e.g. full routing from zeroth order on top to reflection in $-1$ order is denoted $(D\tilde{R}TDDD)$, where channels are numbered clockwise from left top to bottom.  This is expected to require at least ten tuning parameters to be optimally achieved and indeed this is what we find in Fig. 6. The design is optimized by minimizing the smallest singular value of the relevant submatrix of the scattering matrix. As seen in Fig 6, we find that somewhat less suppression is achieved than in the chaotic cavity models, but all the FOMs studied
``dive" down below $20 {\rm dB}$ of suppression at the predicted parameter values, and in this case saturate around $60 {\rm dB}$ of suppression. 
Note also that in this lossless model we again see the dramatic effect of time-reversal symmetry, where perfect reflection from the $-1$ order back into the $+1$ order $(NDDDDD)$ is achieved with only $n_D-n_N + 1=5$ tuning parameters, instead of the ten required when TR symmetry is broken, whereas reflectionless routing into the $-1$ order from the $0$ order (see schematic) requires the full ten. The success of our theory for this structure, with only Fresnel scattering from flat surfaces, suggests that ray chaos is not essential for our codimension argument to be predictive of performance.

In this case the tuning parameters are not geometric, and would be more difficult to reprogram {\em in situ}, but there are a number of methods developed to electrically or optically tune dielectric constants, which might eventually be employed in a similar structure~\cite{bromberg_fiber_piano,Onodera2026}.

\begin{figure}[htp]
    \centering
    \includegraphics[width=0.6\textwidth]{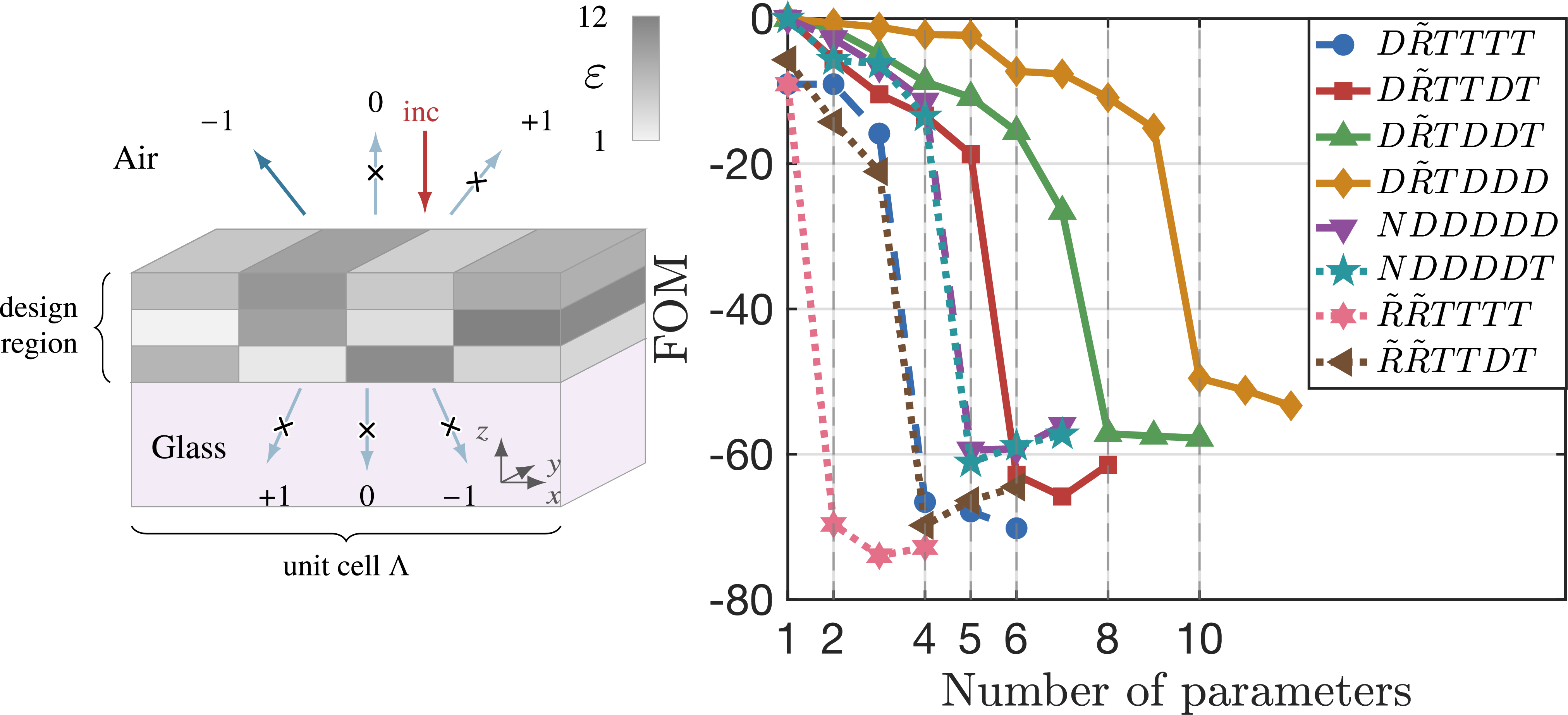}
    \caption{Optimized performance of a periodic dielectric grating (inset) with piecewise-constant permittivity segments as design/tuning parameters. Each curve corresponds to a different CPE process which suppresses reflection and/or transmission diffraction orders for normal incidence, or oblique ($-1$ order) incidence, or coherent excitation from orders $\{-1,0\}$. In every case, the figure of merit saturates once the number of tuning segments reaches that predicted by our codimension arguments. The numerical data in this figure were generated using the rigorous coupled wave analysis (RCWA) package RETICOLO~\cite{reticolo}.}
\end{figure}
\section{Conclusions}

We have developed a novel mathematical framework based on topologically protected complex spectra and codimension arguments to characterize the number of parameters required to control the scattering of monochromatic/energetic waves in  higher dimensional and multiple-scattering geometries.  The approach applies to all linear scattering processes of both classical and quantum matter waves. In contrast to current methods for scattering control, based on black-box optimization and applied extensively to classical waves, this framework is able to characterize the existence of exact solutions via complex spectra and their coincidences on the real axis.  This implies that certain optimization targets of interest for performing information processing functions can be achieved with a relatively small number of design parameters, which in many cases can be tunable {\em in situ}. We showed in extensive simulations that the minimal parameter set predicted by theory can be achieved in plausible models with or without absorption loss and with or without reciprocity. The approach relies on multipath interference and requires a scattering system which is at least a few wavelengths in linear scale.   

Here, for definiteness, we focused on processes which are related to routing and demultiplexing, and are easily extended to mode conversion, isolation and circulation~\cite{yuan2026synthesizing}, a subset of the processes denoted as Coherent Perfect Extinction (CPE). In work in progress we are extending the approach to power distribution and other functions which are not purely CPE, but impose other types of constraints on the S-matrix outputs.

In general, scattering control and functionalization can be achieved in a static design, or in a setup fabricated or pretuned to optimal parameter values, but the approach is particularly attractive when {\em in situ} tuning parameters are available.  In this case arbitrarily reprogrammable performance can be achieved, as has been demonstrated already in the microwave frequency range~\cite{Sol2023ReflectionlessRouters, Faul2025AgileMetasurface}.  There it was shown, in a highly parameterized setup, that narrowband functions could be programmed to operate at any frequency over an enormous frequency band, and reprogrammed to change the detailed routing task or the chosen frequency (or to change the task entirely).

This capability suggests that devices based on these principles would be robust to many environmental perturbations, that simply detune the parameter settings, which could then be retuned optimally to restore the desired function. Such robustness could even be designed to function autonomously. This work opens up many new directions and questions for further research.  {\em In situ} tuning of wave systems has been a major target of research in metamaterials and in integrated optics and photonics generally. One can imagine MEMs and optically modulated systems which eventually will provide new platforms for implementing technologies suggested by this approach.  

Finally, due to the universality of the framework developed here, we are optimistic that it will eventually find applications in design of quantum control and information processing systems and in other areas of wave physics. 

\section*{Acknowledgments}
We are grateful to Hanwen Zhang, Hao He and Ali Ghorashi for helpful discussions and insightful comments. This work was supported by the Simons Collaboration on Extreme Wave Phenomena Based on Symmetries (award no. SFI-MPS-EWP-00008530-09 to A.D.S and O.D.M) and by AFOSR (grant no. FA9550-22-1-0393 to O.D.M).

\appendix

\section{Existence and Density of CCON spectra from TCMT} \label{Existence of CCONs proof}
Starting from Temporal Coupled Mode Theory (TCMT)~\cite{Fan2003TemporalResonators, Suh2004TemporalCavities}, the scattering matrix for a lossless reciprocal system can be written in the compact form
\begin{align}\label{Scattering matrix TCMT}
    S(\omega) = S_0 - i K \frac{1}{\omega - \Omega + i\Gamma}K^\dagger S_0, 
\end{align}
where $K$ describes the coupling from the open channels to the resonant modes, $\Omega$ is a Hermitian matrix of the resonance frequencies (which, with a suitable choice of basis, may be taken to be diagonal) and $\Gamma = K^\dagger K/2$ represents the decay rates.  

Let $F$ and $\bar F$ denote the filtering matrices that select, respectively, the constrained outgoing channels ($D$ and $\tilde{R}$) and the allowed incoming channels ($N$ and $\tilde{R}$). The corresponding constraint matrix is $\mathcal{C} = FS\bar F^\dagger$.  
Applying these filtering matrices to 
~\autoref{Scattering matrix TCMT} yields
\begin{align}
    \mathcal{C}(\omega) = FS_0 \bar F^\dagger - i F K \frac{1}{\omega - \Omega +i\Gamma} K^\dagger S_0 \bar F^\dagger.
\end{align}
We define the filtered coupling matrices for incoming channels $K_F$ and $\tilde{K}_{\bar F}$ for outgoing channels, as
\begin{align}
    K_{F} = FK, \qquad \tilde 
    K_{\bar F} = \bar{F} S_0 K.
\end{align}
$K_F$ here denotes the coupling matrix from the channels in the new basis of (Dark + Reflectionless) to the resonances, and analogously $\tilde{K}_{\bar F}$ represents the coupling matrix restricted to the incoming channels (Normal + Reflectionless). We can also define $\mathcal{C}_0 = FS_0 \bar F^\dagger$ as the filtered background matrix. With the new definitions, we can write:
\begin{align}
    \mathcal{C}(\omega) = \mathcal{C}_0 - i K_F \frac{1}{\omega - \Omega +i\Gamma} \tilde K_{\bar F}^\dagger.
\end{align}
Taking the determinant of both sides and using Schur's determinant formula leaves us with the following

\begin{align}\label{det(C)}
    \det(\mathcal{C}(\omega)) = \det(\mathcal{C}_0)\frac{\det\left(\omega - \Omega +i\Gamma - i\tilde K_{\bar F}^\dagger \mathcal{C}_0^{-1} K_F\right)}{\det\left(\omega - \Omega +i\Gamma \right)}.
\end{align}
So the solutions satisfying the CCON boundary conditions are the eigenvalues of the operator $\Omega -i\Gamma + i\tilde K_{\bar F}^\dagger \mathcal{C}_0^{-1} K_F$. The first two terms of this operator describe the effective Hamiltonian operator of the open system, and the last term corresponds to the shift of CCON zeros from the resonances. This establishes the one to one correspondence between resonances and CCON zeros. Note that if the resonances couple weakly to the incoming channels or the nulled outgoing channels, then the shift is also small (provided that $\mathcal{C}_0^{-1}$ is finite) $\sim \frac{\langle L\mid (\tilde K_{\bar F}^\dagger \mathcal{C}_0^{-1} K_F)\mid R\rangle}{\langle L\mid R\rangle}$,  where $\mid L\rangle$ and $\mid R\rangle$ are the left and right eigenvectors of the effective Hamiltonian operator. 

In the general case, with loss and nonreciprocal elements, the scattering matrix can be written as:
\begin{align}
    S(\omega) = S_0 + i D \frac{1}{\omega - \Omega + i\Gamma}K^T,  
\end{align}
where $D$ and $K$ are not necessarily the same since we don't assume reciprocity. Using a similar procedure we arrive the at the following expression for determinant of the constraint matrix:
\begin{align}
    \det(\mathcal{C}(\omega)) = \det(\mathcal{C}_0)\frac{\det\left(\omega - \Omega +i\Gamma + iK_{\bar F}^T \mathcal{C}_0^{-1} D_F\right)}{\det\left(\omega - \Omega +i\Gamma \right)},
\end{align}
In a lossless, reciprocal system, we have the constraints $D = K$, $S_0 K^* = -K$~\cite{Suh2004TemporalCavities}, which with the fact that the background matrix is unitary and symmetric reduce the last expression to~\autoref{det(C)}.

\putbib 

\end{bibunit}

\clearpage

\newcommand{\highlight}[2]{\colorbox{#1}
{$\displaystyle #2$}}

\setcounter{figure}{0}
\setcounter{table}{0}
\setcounter{equation}{0}

\renewcommand{\thefigure}{S\arabic{figure}}
\renewcommand{\thetable}{S\arabic{table}}
\renewcommand{\theequation}{S\arabic{equation}}

\renewcommand{\figureautorefname}{FIG.}
\renewcommand{\equationautorefname}{Eq.}








\clearpage

\title{Supplementary Material for Coherent Control of Wave Scattering via \\Minimal-Parameter Tuning of Complex Spectra}

\author{Ali H. Alhulaymi}
\email{ali.alhulaymi@yale.edu}
\affiliation{Department of Applied Physics, Yale University, New Haven, CT 06520, USA}

\author{Nazar Pyvovar}
\affiliation{Department of Applied Physics, Yale University, New Haven, CT 06520, USA}

\author{Philipp del Hougne}
\affiliation{Univ Rennes, CNRS, IETR-UMR 6164, F-35000 Rennes, France}

\author{Owen D. Miller}
\affiliation{Department of Applied Physics, Yale University, New Haven, CT 06520, USA}

\author{A. Douglas Stone}
\email{douglas.stone@yale.edu}
\affiliation{Department of Applied Physics, Yale University, New Haven, CT 06520, USA}

\maketitle
\phantomsection
\label{title:SM}

\addtocontents{toc}{\protect\showSupplementTOCEntries}

\tableofcontents

\setcounter{section}{0}
\setcounter{subsection}{0}
\setcounter{subsubsection}{0}

\section{CCONs as Eigenvalues of a Generalized Operator}
In this section, we introduce the channel boundary conditions notation and derive the eigenvalue equation within Temporal Coupled Mode Theory (TCMT). 
\subsection{Schur's Formula}
We start by providing a proof of Schur's determinant formula \cite{Horn2013MatrixAnalysis, Meyer2000MatrixAlgebra}. Let 

\begin{align}
    M = \begin{pmatrix} A & B \\ C & D\end{pmatrix},
\end{align}
where $A$ and $D$ are square matrices. When $A$ is invertible, the determinant of $M$ satisfies

\begin{align}\label{Schur formula 1}
   \det(M) = \det(A) \det(D - CA^{-1}B),
\end{align}
 and when $D$ is invertible, we have
 
\begin{align}\label{Schur formula 2}
   \det(M) = \det(D) \det(A - BD^{-1}C). 
\end{align}
As in \cite{Meyer2000MatrixAlgebra}, this is proven by noting that 
\begin{align}
    \begin{pmatrix} A & B \\ C & D\end{pmatrix} = \begin{pmatrix} I & 0 \\ CA^{-1} & I\end{pmatrix} \begin{pmatrix} A & B \\ 0 & D - CA^{-1}B\end{pmatrix}.
\end{align}
Taking the determinant of both sides leads to~\autoref{Schur formula 1}. The second form~\autoref{Schur formula 2} is proven in an analogous manner.

\subsection{The constraint matrix}\label{The constraint matrix definition}
For linear systems, the scattering matrix relates the incoming and outgoing channel amplitudes via

\begin{align}
    \beta(\omega) = S(\omega) \alpha,
\end{align}
where $\alpha$ and $\beta$ are the vectors of coefficients in the incoming and outgoing channel bases respectively.
Let $\mathcal{I} = \{1, \cdots , N_c\}$ be the index set of all channels. We can define the following subsets of $\mathcal{I}$, according to the conditions imposed on $\alpha$ and $\beta$.

\begin{align}
    \mathcal{I}_{\tilde{R}} &= \{i \in \mathcal{I} \mid \beta_i = 0\}, &\qquad \mathcal{I}_{T} &= \{i\in \mathcal{I} \mid \alpha_i = 0 \} \nonumber\\ 
    \mathcal{I}_{D} &= \{i\in \mathcal{I}\mid \alpha_i = 0 \; \text{and}\; \beta_i = 0\}, &\qquad \mathcal{I}_{N} &= \mathcal{I}\backslash(\mathcal{I}_{\tilde{R}}  \cap \mathcal{I}_T)  \nonumber
\end{align}
Note that with this definition 
$\mathcal{I}_D = \mathcal{I}_{\tilde{R}} \cap \mathcal{I}_T$. We also define the incoming and zero-outgoing index sets as

\begin{align}
    \mathcal{I}_{in} = \mathcal{I}_{\tilde{R}} \cup \mathcal{I}_N, \qquad \mathcal{I}_{0} = \mathcal{I}_{\tilde{R}} \cup \mathcal{I}_D
\end{align}
For a given set of boundary conditions with $n_{\tilde{R}}$ reflectionless channels, $n_T$ transmission channels, $n_N$ normal channels and $n_D$ dark channels, the relevant submatrix whose determinant is zero at the solutions to the corresponding boundary conditions will be denoted by $\mathcal{C}$. The constraint matrix is an $m\times n$ matrix, with $m = n_D + n_{\tilde{R}}$ and $n = n_N + n_{\tilde{R}}$ . 
This submatrix is obtained from the full scattering matrix by use of filtering matrices $F$ and $\bar F$ defined by 
\begin{align}
    F = (e_{i_1}, e_{i_2}, \cdots, e_{i_n})^\dagger,
\end{align}
where $i_r \in \mathcal{I}_{in}$ is in the set of indices of channels with either a normal or a reflectionless constraint, $e_s$ is the $N$ dimensional basis vector $(0, 0, \cdots, 1, \cdots, 0)^T$ with 1 at $s$ and zero for the remaining entries. For $\bar{F}$ we have 
\begin{align}
    \bar{F} = (e_{j_1}, e_{j_2}, \cdots, e_{j_m})^\dagger,
\end{align}
where $j_r \in \mathcal{I}_0$ is in the set of indices of channels with either a reflectionless or a dark channel. 
From this, the constraint matrix, could be written as
\begin{align}
    \mathcal{C} = F S \bar F^\dagger
    \label{C_definition}
\end{align}
An example is below
\begin{align}
    \mathcal{C} = \begin{pNiceMatrix}
        S_{11} & \cdots & S_{1i} &  \cdots & S_{1j} &\cdots & S_{1n}\\
        \vdots & \ddots & \vdots & \ddots & \vdots & \ddots & \vdots\\
        S_{k1} & \cdots & S_{ki} & \cdots & S_{kj}& \cdots & S_{kn}\\
        \vdots & \ddots & \vdots & \ddots & \vdots & \ddots &\vdots\\
        S_{l1} & \cdots & S_{li} & \cdots & S_{lj}& \cdots & S_{ln}\\
        \vdots & \ddots & \vdots & \ddots & \vdots & \ddots & \vdots \\
        S_{m1} & \cdots & S_{mi} &  \cdots & S_{mj} &\cdots & S_{mn}\\
    \CodeAfter
    \tikz \node[draw=myblue, dashed, line width = 1pt, fit=(3-1)(3-7), inner sep=2pt] {};
    \tikz \node[draw=myblue, dashed, line width = 1pt, fit=(5-1)(5-7), inner sep=2pt] {};
    \tikz \node[draw=myred, dashed, line width = 1pt, fit=(1-3)(7-3), inner sep=2pt] {};
    \tikz \node[draw=myred, dashed, line width = 1pt, fit=(1-5)(7-5), inner sep=2pt] {};
        \end{pNiceMatrix}.
\end{align}
At a given frequency $\omega_0$, a nontrivial solution that satisfies the boundary conditions exists if and only if the associated constraint matrix has a nontrivial nullity. For CCONs, $n_D = n_N$ so the constraint matrix is square,  and the condition reduces to $\det(\mathcal{C}(\omega_0)) = 0$. For all other boundary conditions, the constraint matrix is rectangular: when $n_D<n_N$ the system is underdetermined (a trivial case which we do not consider here), whereas the problem with $n_D>n_N$ gives an overconstrained problem. The special case with $m=n=0$, is the all $T$ class, which is the purely outgoing resonance condition, defined by a pole of the scattering matrix. 

\subsection{TCMT formula for CCON Spectra}
Starting from Temporal Coupled Mode Theory (TCMT)~\cite{Fan2003TemporalResonators, Suh2004TemporalCavities}, the scattering matrix for a lossless reciprocal system can be written in the compact form
\begin{align}\label{Scattering matrix TCMT}
    S(\omega) = S_0 - i K \frac{1}{\omega - \Omega + i\Gamma}K^\dagger S_0,  
\end{align}
where $K$ describes the coupling from the open channels to the resonant modes, $\Omega$ is a Hermitian matrix of the resonance frequencies (which, with a suitable choice of basis, may be taken to be diagonal) and $\Gamma = K^\dagger K/2$ represents the decay rates. 
Applying the filtering matrices to~\autoref{Scattering matrix TCMT} yields
\begin{align}
    \mathcal{C}(\omega) = FS_0 \bar F^\dagger - i F K \frac{1}{\omega - \Omega +i\Gamma} K^\dagger S_0 \bar F^\dagger.
\end{align}
We define the filtered coupling matrices for incoming channels $K_F$ and $\tilde{K}_{\bar F}$ for outgoing channels, as
\begin{align}
    K_{F} = FK, \qquad \tilde 
    K_{\bar F} = \bar{F} S_0 K.
\end{align}
$K_F$ here denotes the coupling matrix from the channels in the new basis of (Dark + Reflectionless) to the resonances, and analogously $\tilde{K}_{\bar F}$ represents the coupling matrix restricted to the incoming channels (Normal + Reflectionless). We can also define $\mathcal{C}_0 = FS_0 \bar F^\dagger$ as the filtered background matrix. With the new definitions, we can write:
\begin{align}
    \mathcal{C}(\omega) = \mathcal{C}_0 - i K_F \frac{1}{\omega - \Omega +i\Gamma} \tilde K_{\bar F}^\dagger.
\end{align}
Taking the determinant of both sides and using Schur's determinant formula leaves us with the following

\begin{align}\label{det(C)}
    \det(\mathcal{C}(\omega)) = \det(\mathcal{C}_0)\frac{\det\left(\omega - \Omega +i\Gamma - i\tilde K_{\bar F}^\dagger \mathcal{C}_0^{-1} K_F\right)}{\det\left(\omega - \Omega +i\Gamma \right)}.
\end{align}
So the solutions satisfying the CCON boundary conditions are the eigenvalues of the operator $\Omega -i\Gamma + i\tilde K_{\bar F}^\dagger \mathcal{C}_0^{-1} K_F$. The first two terms of this operator describe the effective Hamiltonian operator of the open system, and the last term corresponds to the shift of CCON zeros from the resonances. Note that if the resonances couple weakly to the incoming channels or the nulled outgoing channels, then the shift is also small (provided that $\mathcal{C}_0^{-1}$ is finite) $\sim \frac{\langle L\mid (\tilde K_{\bar F}^\dagger \mathcal{C}_0^{-1} K_F)\mid R\rangle}{\langle L\mid R\rangle}$,  where $\mid L\rangle$ and $\mid R\rangle$ are the left and right eigenvectors of the effective Hamiltonian operator. 

In the general case, with loss and nonreciprocal elements. The scattering matrix can be written as:
\begin{align}
    S(\omega) = S_0 + i D \frac{1}{\omega - \Omega + i\Gamma}K^T,  
\end{align}
where $D$ and $K$ are not necessarily the same since we don't assume reciprocity. Using a similar procedure we arrive the at the following expression for determinant of the constraint matrix:
\begin{align}
    \det(\mathcal{C}(\omega)) = \det(\mathcal{C}_0)\frac{\det\left(\omega - \Omega +i\Gamma + iK_{\bar F}^T \mathcal{C}_0^{-1} D_F\right)}{\det\left(\omega - \Omega +i\Gamma \right)}.
\end{align}
In a lossless, reciprocal system, we have the constraints $D = K$, $S_0 K^* = -K$~\cite{Suh2004TemporalCavities}, which with the fact that the background matrix is unitary and symmetric reduce the last expression to~\autoref{det(C)}.

\subsection{$S_0 = \mathbb{I}_{N_c}$ case} \label{S_0 = I}

When the background matrix is the identity, a problem arises in~\autoref{det(C)}. For cases where there is at least 1 dark channel, the filtered background matrix $\mathcal{C}_0$ is singular and we can no longer apply Schur's determinant formula. However, there is a nice way to rewrite the expression for the determinant of the constraint matrix. We first relabel our channels so that we can write the CCON as $(N\cdots N, D\cdots D\tilde{R}\cdots \tilde{R} T\cdots T)$. With this choice, we can write
\begin{align}
    \mathcal{C}(\omega) = \begin{pmatrix}
        0_{n_D\times n_N} & 0_{n_D\times n_{\tilde{R}}}\\
        0_{n_{\tilde{R}}\times n_N} & \mathbb{I}_{n_{\tilde{R}}\times n_{\tilde{R}}} \end{pmatrix} - i \begin{pmatrix}
            K_D\\K_{\tilde{R}}
        \end{pmatrix}G ^{-1}(\omega)\begin{pmatrix}
            K_N^\dagger & K_{\tilde{R}}^\dagger
        \end{pmatrix}.
\end{align}
where $G(\omega) = \omega - \Omega +i\Gamma$.  
Taking the determinant of both sides and using the second form of Schur's determinant formula gives 
\begin{align}
    \det(\mathcal{C}(\omega)) =\frac{1}{\det(G(\omega))} \det\begin{pmatrix}
        0_{n_D\times n_N} & 0_{n_D\times n_{\tilde{R}}} & i K_D\\
        0_{n_{\tilde{R}}\times n_N} & \mathbb{I}_{n_{\tilde{R}}\times n_{\tilde{R}}} & iK_{\tilde{R}}\\
        K_N^\dagger & K_{\tilde{R}}^\dagger & G(\omega)\\
    \end{pmatrix}. 
\end{align}
Using the same formula again but with $A = 0_{n_D\times n_N}$, $B = \begin{pmatrix}
    0_{n_D\times n_{\tilde{R}}}, & i K_D
\end{pmatrix}$,  $C = \begin{pmatrix} 0_{n_N \times n_{\tilde{R}}},& K_N
\end{pmatrix}^\dagger$ and $D$ is the remaining block matrix, gives 
\begin{align}
    \det(\mathcal{C}(\omega)) = \frac{1}{\det(G(\omega))}\det \begin{pmatrix}
        \mathbb{I}_{n_{\tilde{R}}\times n_{\tilde{R}}} &iK_{\tilde{R}}\\
        K_{\tilde{R}}^T & G(\omega)
    \end{pmatrix} \det\left[-\begin{pmatrix}
    0_{n_D\times n_{\tilde{R}}} & i K_D
\end{pmatrix}\begin{pmatrix}
        \mathbb{I}_{n_{\tilde{R}}\times n_{\tilde{R}}} &iK_{\tilde{R}}\\
        K_{\tilde{R}}^T & G(\omega)
    \end{pmatrix}^{-1}\begin{pmatrix}
        0_{n_{\tilde{R}}\times n_N} \\ K_N^\dagger
    \end{pmatrix}\right]
\end{align}
We can simplify the first term using Schur's determinant formula and the second factor using the inverse formula. 
\begin{align}
    \det(\mathcal{C}(\omega)) = \frac{\det(G(\omega) - iK_{\tilde{R}}^\dagger K_{\tilde{R}})\det\left(-iK_D\left(G(\omega) - iK_{\tilde{R}}^\dagger K_{\tilde{R}}\right)^{-1} K_N^\dagger\right)}{\det G(\omega)}
\end{align}
With $G(\omega) = H_0(\omega) +  i\Gamma$, for any $H_0(\omega)$, as long as $S_0 = \mathbb{I}_{N_c}$ (e.g., Quantum Graphs, Heidelberg model, TCMT):
\begin{align}
    \det(\mathcal{C}(\omega)) = \frac{\det \left(H_0(\omega) + i \Gamma_{\rm eff} \right)\det\left(-iK_D(H_0(\omega) +i\Gamma_{\rm eff})^{-1} K_N^\dagger\right)}{\det G(\omega)},
\end{align}
where $\Gamma_{\rm eff} = \Gamma_N + \Gamma_D + \Gamma_T- \Gamma_{\tilde{R}}$, is the effective decay rate, and $\Gamma_C = K_C^\dagger K_C/2$. 
The interpretation of the last equation is that the $\tilde{R}$ channels in CCONs always serve as an effective gain. 
For $H_0(\omega) =\omega -\Omega $ (i.e.,  TCMT), we can multiply the right hand side by $\frac{\omega^{n_D}}{\omega^{n_D}}$ to obtain
\begin{align} \label{det C final form}
    \det(\mathcal{C}(\omega)) = \frac{\det \left(-iK_D K_N^\dagger \right)\det \left(\omega - \left(\mathbb{I}_{N_{\rm res}} - K_N^\dagger(K_D K_N^\dagger)^{-1}K_D \right)(\Omega - i\Gamma_T + i\Gamma_{\tilde{R}})\right)}{\omega^{n_D} \det\left(\omega - \Omega +i\Gamma \right)}.
\end{align}
So the CCON solutions are the non-zero eigenvalues of the operator $\left(\mathbb{I}_{N_{\rm res}} - K_N^\dagger(K_D K_N^\dagger)^{-1}K_D\right)(\Omega - i\Gamma_T + i\Gamma_{\tilde{R}})$. Here we exclude exactly $n_D$ zero eigenvalues that were introduced because we multiplied the numerator and denominator by $\omega^{n_D}$. This explicitly shows that when the background is ignored, we lose $n_D$ solutions that are pushed off to infinity. 

\section{Random Matrix Theory}
\begin{figure}[H]
    \centering  \includegraphics[width=0.8\textwidth]{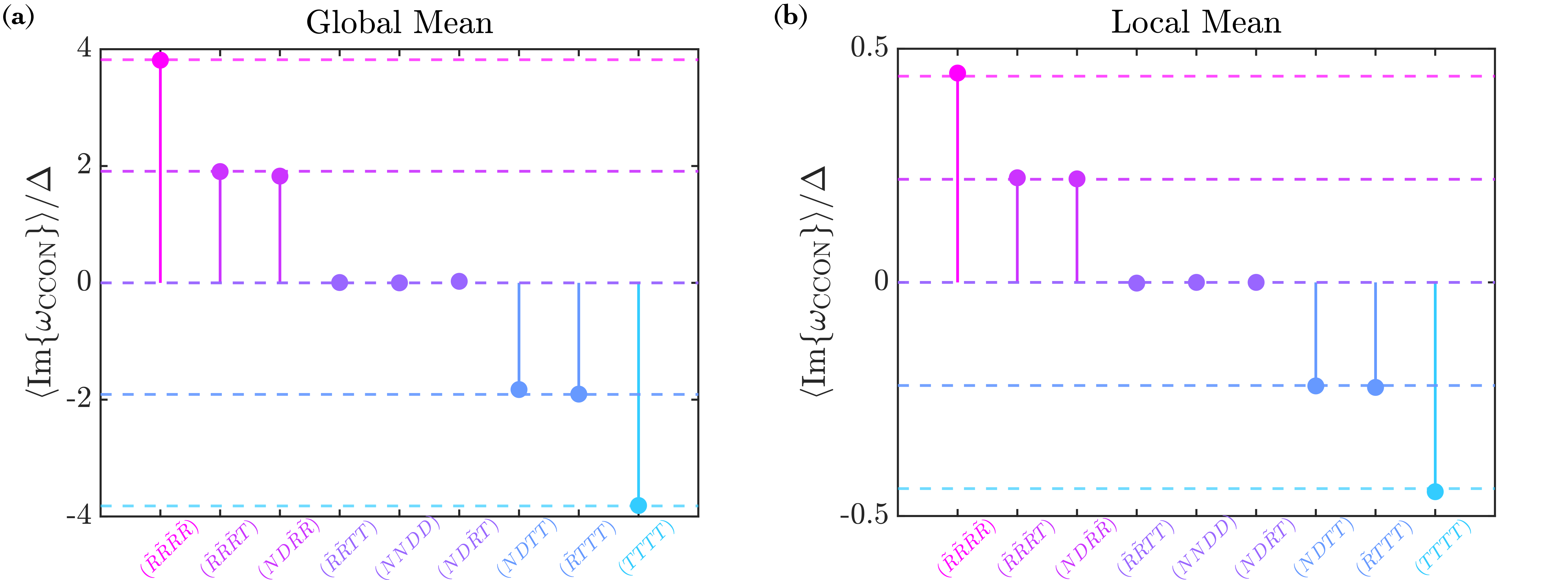}
    \caption{Numerical results for the (a) global and (b) local mean of the imaginary parts of CCONs. Each point is computed over an ensemble of 500 realizations and $N_{\rm res} = 400$ resonances each. Parameters: $\gamma=1.5$ ,  $\lambda = 1/2$. The dashed lines show the predicted averages for CCONs with $n_{\tilde{R}} - n_T = (4, 2, 0, -2, -4)$. The numerical results show excellent agreement with the theoretical predictions. When estimating the local mean numerically, superradiant-like states were excluded from the averaging. }
    \label{fig:Imaginary part of CCONs}
\end{figure}

In Random Matrix Theory (RMT), the scattering matrix has the same form as in TCMT:
\begin{align}
    S(\omega) = \mathbb{I} - i W^\dagger \frac{1}{\omega - H + i\Gamma}W ,  
\end{align}
where $H$ is the closed cavity Hamiltonian, and $W_{\alpha i}$ is the coupling amplitude from the $\alpha$ resonance to the $i_{th}$ open channel. To obtain the statistical average of the imaginary part of CCON eigenfrequencies, we adopt the conventional way of taking $H$ from a GOE ensemble \cite{Mitchell2010RandomReactions, Fyodorov2015ResonancePorter-Thomas} with
\begin{align}
    \langle H \rangle = 0, \qquad \langle H_{\alpha \beta}H_{\alpha' \beta'}\rangle = \frac{\lambda^2}{N_{\rm res}}\left(\delta_{\alpha \alpha'} \delta_{\beta \beta'} + \delta_{\alpha \beta} \delta_{\alpha' \beta'} \right).
\end{align}
With the mean level spacing around the center is $\Delta = \pi \lambda/N_{\rm res}$.
We also take each element of the coupling matrix $W$ from a normal distribution such that 
\begin{align}
    \langle W_{\alpha i} \rangle = 0, \qquad \langle W^*_{\alpha i} W_{\beta j} \rangle = 2\frac{\gamma_\alpha}{N_{\rm res}} \delta_{\alpha \beta} \delta_{ij}. 
\end{align}
To obtain a global statistical average, we use \autoref{det C final form}, where the CCON solutions are the eigenvalues of $\\ \left(\mathbb{I}_{N_{\rm res}} - W_N(W_D^\dagger W_N)^{-1}W_D^\dagger\right)(H - i\Gamma_T + i\Gamma_{\tilde{R}})$. Here $W_C$ and $\Gamma_C$ are defined similarly to \autoref{S_0 = I}.
The global average will then be the ensemble average over all eigenvalues in all realizations, that is 
\begin{align}
    \langle \omega_{ \rm CCON}\rangle =\frac{1}{N_{\rm res} - n_D} \left\langle \Tr{\left(\mathbb{I}_{N_{\rm res}} - W_N(W_D^\dagger W_N)^{-1}W_D^\dagger\right)(H - i\Gamma_T + i\Gamma_{\tilde{R}})} \right\rangle,
\end{align}
where the brackets refer to the ensemble average. Since the two factors are statistically independent, the average can be obtained by evaluating the mean of each factor and then multiplying. So we can start with the second factor first. 
\begin{align}
\langle H - i\Gamma_T + i\Gamma_{\tilde{R}} \rangle &= 0 - i\langle W_TW_T^\dagger\rangle/2 + i\langle W_{\tilde{R}} W_{\tilde{R}}^\dagger \rangle /2
\\ &= -i  n_T \frac{\gamma}{N_{\rm res}}\mathbb{I}_{N_{\rm res}} + i   n_{\tilde{R}} \frac{\gamma}{N_{\rm res}} \mathbb{I}_{N_{\rm res}}\\
&= i(n_{\tilde{R}} - n_T) \frac{\gamma}{N_{\rm res}} \mathbb{I}_{N_{\rm res}}.
\end{align}
With this, we can write
\begin{align}
    \langle \omega_{ \rm CCON}\rangle = i(n_{\tilde{R}} - n_T) \frac{\gamma}{N_{\rm res}(N_{\rm res} - n_D) } \left(N_{\rm res} - \left\langle \Tr{W_N(W_D^\dagger W_N)^{-1}W_D^\dagger} \right\rangle \right).
\end{align}
Simplifying the last expression leaves us with
\begin{align}
    \langle \omega_{ \rm CCON}\rangle = i(n_{\tilde{R}} - n_T) \frac{\gamma}{N_{\rm res}}.
\end{align}
The expression given above refers to the \emph{global mean} or the Weisskopf estimate~\cite{Mitchell2010RandomReactions, Fyodorov2015ResonancePorter-Thomas}, which differs from the \emph{local mean}. The global mean obtained via the trace formula, is valid in the weak-coupling regime. By contrast,  the local mean of resonance widths at finite coupling is estimated by the Moldauer-Simonius (MS) relation~\cite{Fyodorov2015ResonancePorter-Thomas, Mitchell2010RandomReactions}. Intuitively, the mean imaginary part should have a weak dependence on  $n_D$ since in the operator $\left(\mathbb{I}_{N_{\rm res}} - W_N(W_D^\dagger W_N)^{-1}W_D^\dagger\right)(H - i\Gamma_T + i\Gamma_{\tilde{R}})$, the factor that carries the dependence on both dark and normal channels multiplies the Hermitian and anti-Hermitian parts of the operator in the same way, therefore, its effect should be weak on the mean of the imaginary part of CCONs when normalized by the mean level spacing $\Delta$. Motivated by this, we expect an MS-type relation to hold for CCONs:
\begin{align}
    \frac{\langle {\rm Im}\{\omega_{\rm CCON}\}\rangle}{\Delta} = -\frac{(n_{\tilde{R}} - n_T)}{4\pi} \ln(1- T),
\end{align}
where we have assumed equivalent channels with $T = T_c = \frac{4\kappa_c}{(1+\kappa_c)^2}$, and $\kappa_c = \frac{\pi \gamma_c}{N_{\rm res} \Delta}$.
Thus, $ND$ pairs are expected not to contribute strongly to the average displacement of CCONs from the real axis. A detailed RMT analysis of CCON statistics will appear in a future publication.~\autoref{fig:Imaginary part of CCONs} Shows a numerical demonstration of the independence of the global and local mean on $ND$  pairs, and showing excellent agreement with our predictions.
\subsection{Special Symmetry of Perfect Reflection CCON}

We have proved the existence and discreteness of all CCON spectra in the Appendix based on the properties of the meromorphic function $det(C)$ and we have constructed approximate formulas for $det(C)$ within TCMT or equivalently from the Heidelberg RMT representation. For a general CCON those representations show no special properties or symmetries of CCON spectra other than their discrete complex nature and their one-to-one correspondence with the resonances of the system. 

However when the system is reciprocal and lossless, so S is a symmetric unitary matrix, there is one exception, balanced perfect reflection in a system with an even number of ports. This is a CCON with exactly $N_c2$ normal ports and $N_c/2$ dark ports. Looking at such processes is particularly natural when the system of interest is quasi-1D and statistically identical on the ``left" and``right" side.  Choosing the left side as input (normal) ports and the right as the dark ports defines this CCON as a zero of the $N_c/2 \times N_c/2$ transmission matrix, $T$.  In a study by Kang et al.~\cite{Kang2021TransmissionSystems}, prior to our current work on CCONs, the authors considered a quasi-1D disordered or chaotic scattering system of this form, both in RMT and numerically. They were able to show from analyzing the RMT expression for the transmission matrix that for such a matrix the transmission zeros come in complex conjugate pairs or are real and isolated (generically).  Note that the same property characterizes the eigenfrequencies (or energies) of non-Hermitian systems with parity-time symmetry, but in that case the scattering system must satisfy spatial symmetries as well, which isn't required here.  Note that in the complex plane the transmission zeros are still discrete in this case and have codimension two, like all other CCONs.  

However, as mentioned in the main text, this property underlies our discovery that the constraint $\det(T)=0$ has {\it codimension one} in a generic real parameter space (where here we adopt $T$ as our notation for the relevant constraint matrix), meaning that it can be satisfied at a fixed frequency by tuning a single efficient parameter. 
Here we show briefly that the property of the balance perfect reflection CCON spectrum follows exactly from the time-reversal symmetry of unitary reciprocal S-matrices without resorting to an approximate analytic representation.  

For a lossless reciprocal scattering system if $S{\bf a}= {\bf b}$, it follows from time-reversal that $S(\omega^*){\bf b^* }= {\bf a}^*$. For the perfect reflection CCON, the output vector ${\bf b}$ is only non-zero in the outgoing subspace corresponding to the time-reverse of the $N_c/2$ filled input channels. Hence the scattering process at $\omega^*$ is also a perfect reflection CCON, for the same set of ports. Thus every such CCON must either come in complex conjugate pairs or occur on the real axis.

\section{Implementing CCON boundary conditions as eigenvalue problems}
The purpose of this section is to recast the CCON problem as a finite-element eigenvalue problem. The central idea is to treat the incident amplitudes in the $N$ and $\tilde{R}$ channels as additional unknowns, while the vanishing outgoing amplitudes in $D$ and $\tilde {R}$ are imposed as constraints. The resulting problem could be solved numerically after discretizations. This eigenvalue formulation is a generalization of the approach proposed in~\cite{Shao1995AnProblems} to systems with more than two ports and to higher-dimensional settings.

\subsection{Weak formulation of the Helmholtz equation}
Assuming no sources are present, Maxwell equation in 2D reduces to the Helmholtz equation for the scalar field $u = H_z$:
\begin{align}
    \nabla^2 u  + n^2 k^2 u = 0. 
\end{align}

We can multiply the last equation by a test function $v$ and integrate over the volume of the scattering region to obtain

\begin{align}
    \int_\Omega (\nabla^2 u  + n^2 k^2 u) v^* \dd\Omega
    = 0. 
\end{align}

Using Green's first identity on the first term,  we get

\begin{align}\label{General weak form}    
    \int_\Omega \left(\nabla u \cdot \nabla v^*  - k^2 n^2 uv^*\right) \dd \Omega - \int_\Gamma \partial_n u v^* \dd \Gamma = 0
\end{align}

Along the ports, we can expand the solution in the channel basis
\begin{align}
    u(s_m, z_m) = \sum_p \left(a_{pm} e^{-i\beta_{pm} z_m} + b_{pm} e^{i\beta_{pm} z_m}\right) \phi_{pm}(s_m),
\end{align}
where $z$ is the propagation direction, $s$ is the transverse direction, $\beta_{pm}$ is the $p_{th}$ propagating mode in the $m_{th}$ channel and $\phi_{pm}$ is the transverse wave profile of the $p_{th}$ propagating mode in the $m_{th}$ channel. $a$ and $b$ denote the coefficients of the incoming and outgoing modes. 

For simplicity, we assume we have rectangular metallic waveguides, in which we have 

\begin{align}
    \beta_{pm} = \sqrt{n_m^2k^2 - \alpha_{pm}^2},
\end{align}

The transverse wave profile has to also satisfy the orthogonality relation
\begin{align}
    \langle\phi_{pm}, \phi_{qm}\rangle = \int_{\Gamma_m} \phi_{pm} \phi^*_{qm} \dd \Gamma= \delta_{pq}.
\end{align}

At the boundary of the waveguides and assuming we keep the relevant propagating and evanescent modes in the expansion, we can write the field and its normal derivative at a boundary along the waveguide as

\begin{align}
    u\big|_{\Gamma_m} &=  \sum_p \left(a_{pm} + b_{pm} \right) \phi_{pm}(s_m),\\
    \frac{\partial}{\partial z_m} u \big|_{\Gamma_m}&= \sum_p i \beta_{pm}\left(-a_{pm} + b_{pm} \right) \phi_{pm}(s_m).
\end{align}
Solving for the coefficients of the outgoing amplitudes gives

\begin{align} \label{Outgoing amplitudes}
    b_{pm} = c_{pm}(u) - a_{pm},
\end{align}
where $c_{pm}(u) = \langle u, \phi_{pm}\rangle_{\Gamma_m}$.

From this we can write
\begin{align}
    \frac{\partial}{\partial z_m} u \big|_{\Gamma_m}&= \sum_p i \beta_{pm}\left(-2 a_{pm} + c_{pm} \right) \phi_{pm}(s_m).
\end{align}

We can plug this back into \autoref{General weak form}, to obtain 

\begin{align}\label{Weak form with incident wave}
    \int_\Omega \left(\nabla u \cdot \nabla v^*  - k^2 n^2 uv^*\right) \dd \Omega - \int_{\Gamma_{\text{wall}}}\partial_n u v^* \dd \Gamma- \sum_{p,m} i\beta_{pm} c_{pm}(u) c_{pm}^*(v) = -\sum_{p,m} 2ia_{pm} c_{pm}^*(v)\beta_{pm} 
\end{align}
This last equation describes the driven scattering problem with the incoming amplitudes playing the role of sources. For many standard FEM packages, it is enough to describe the weak boundary terms as weak form contribution. In COMSOL, for example, the whole problem can be modeled in the PDE module. We briefly describe how to model it there and then in the next section give a more complete description with the discretizations process. 

\subsubsection{Implementation in COMSOL}
Equation \autoref{General weak form} can be implemented directly with a Weak Form PDE interface. Any non-PEC boundary condition can be added as a separate weak contribution. On port $m$, we add a boundary weak contribution whose integral is 
\begin{align}
    \int_{{\Gamma}_m} \left(-\sum_p i\beta_{pm}c_{pm}(u)\phi_{pm}(s_m) + \sum_{p\in\mathcal{I}_{in}} 2i\beta_{pm}a_{pm}\phi_{pm}(s_{m})\right)v^*\dd \Gamma
\end{align}

We also need to introduce one complex global dependent variable $a_{pm}$ for each channel in $\mathcal{I}_{N}$ and the zero-output conditions for each channel in $\mathcal{I}_{D}$.  These constraints are imposed as global equations and as long as $n_D = n_N$ (which is true for all CCONs) this will result in a square eigenvalue problem in the fields and the incident amplitudes. The $\tilde{R}$ and $T$ boundary conditions can be imposed by imposing purely incoming or purely outgoing boundary conditions.

\subsection{Finite element matrix form}
To arrive at the matrix form and the eigenvalue problem to be solved, we start by writing the field and the test function in terms of some shape functions

\begin{align}
    u &= \sum_j f_j U_j \\
    v &= \sum_j g_j U_j,
\end{align}
where $h_i$ and $g_i$ are complex coefficients. This allows to rewrite \autoref{Weak form with incident wave} as

\begin{align}
    \sum_{i, j} f_i g_j^* \int_\Omega \nabla U_i \cdot \nabla U_j^* \dd\Omega  - k^2 \sum_{i, j} \int_\Omega n^2 U_i U_j^* \dd\Omega - \sum_{i, j}f_i g_j^* \int_{\Gamma_{\text{wall}}} \partial_n U_i U_j^* \dd \Gamma -\sum_{p, m} i \beta_{pm} \sum_{i, j} f_i g_j^* c_{pm}(U_i) c_{pm}^*(U_j) \nonumber \\= -\sum_j g_j^* \sum_{p, m} 2ia_{pm} \beta_{pm} c_{pm}^*(U_j) 
\end{align}

We define the following operators 

\begin{align}
    L_{ij} &= \int_\Omega \nabla U_i \cdot \nabla U_j^* \dd\Omega \\
    M_{ij} &= \int_\Omega n^2 U_i U_j^* \dd\Omega
    \\
    G_{ij} &= \int_{\Gamma_{\text{wall}}} \partial_n U_i U_j^* \dd \Gamma
    \\
    C_{ij} &= \sum_{p, m} i\beta_{pm} c_{pm}(U_i) c_{pm}^*(U_j)
    \\
    A_j &= -\sum_{p, m} 2 i\beta_{pm} a_{pm} c_{pm}^*(U_j).
\end{align}
Now we can write

\begin{align}
    \sum_{j} g_j^* \left(\sum_{i} L_{ij}f_i\right) -  \sum_{j} g_j^* k^2\left(\sum_{i} M_{ij}f_i\right)  - \sum_{j} g_j^* \left(\sum_{i} G_{ij}f_i\right) - \sum_{j} g_j^* \left(\sum_{i} C_{ij}f_i\right) &= \sum_{j} g_j^* A_j .  
\end{align}
This equation should valid for all functions and so it is true for all $g_j$. Thus, 
\begin{align}
    \left(\mathbf{L} - \mathbf{G}(k) - \mathbf{C}(k) - k^2 \mathbf{M}  \right)\mathbf{f} = \mathbf{A}
\end{align}

For a fixed the $k$, the driven scattering problem is solved by a simple matrix inversion. 

\subsection{Imposing the CCON boundary conditions}
For the CCON problem, we can impose the boundary conditions by setting the outgoing amplitudes corresponding to $\tilde{R}$ and $D$ channels to zero. We first introduce a column vector for the contribution of each channel to simplify the subsequent expressions:
\begin{align}
    (\mathbf{q}_{pm})_i &= c^*_{pm}(U_i).
\end{align}
It is also convenient to introduce the operator $\mathbf{P}$ as
\begin{align}
    \mathbf{P}_{pm} = -2i\beta_{pm} \mathbf{q}_{pm}
\end{align}

Now, with $\mathcal{I}_{in}$ being the set of channel indices where the incoming  amplitudes are not constrained to zero (see \autoref{The constraint matrix definition}, and $\mathcal{I}_0$ be the set of channel indices where the outgoing amplitudes are set to zero, then 

\begin{align}
    \left(\mathbf{L} - \mathbf{G}(k) - \mathbf{C}(k) - k^2 \mathbf{M}  \right)\mathbf{f} = \mathbf{P}_{\mathcal{I}_{in}}a_{\mathcal{I}_{in}}
\end{align}

Now imposing the constraints for each $(p, m) \in \mathcal{I}_{0}$ gives

\begin{align}
    0 = b_{pm} = \mathbf{q}_{pm}^\dagger \mathbf{f} - a_{pm}
\end{align}

More compactly we can write 

\begin{align}
   \mathbf{q}_{pm}^\dagger \mathbf{f} - \sum_{(q, n) \in \mathcal{I}_{in}} \delta_{pq} \delta_{mn} a_{pm} = 0,
\end{align}
where the Kronecker delta functions account for reflectionless channels in which $a_{pm}$ is nonzero even though the outgoing amplitude is set to zero.

\begin{align} \label{Finite Element eigenvalue problem}
    \begin{pmatrix}
        \mathbf{L} - \mathbf{G}(k) - \mathbf{C}(k) - k^2 \mathbf{M}  && -\mathbf{P}_{\mathcal{I}_{in}}\\
        \mathbf{Q}^\dagger_{\mathcal{I}{0}} && -\mathbf{\Delta}_{\mathcal{I}_0}
    \end{pmatrix}   \begin{pmatrix}
        \mathbf{f} \\ \mathbf{a}_{\mathcal{I}_{in}}
    \end{pmatrix}  = \begin{pmatrix}
        0\\0
    \end{pmatrix}.
\end{align}
This block matrix equation is the CCON eigenvalue problem. Its eigenvector contains both the field solution and the incident wavefront. The block matrix is square whenever $\big| \mathcal{I}_{in}\big| = \big| \mathcal{I}_{0}\big|$ or $n_N = n_D$, the same condition for the general existence of solutions for CCONs. Additionally, because of the wall term $\mathbf{G}$ and the terms containing the propagation constant for each channel $\beta_{pm}$, \autoref{Finite Element eigenvalue problem} is in general a nonlinear eigenvalue problem in $k$, but for most cases it can be written as a quadratic eigenvalue problem.

With Neumann boundary conditions on the walls of the cavity and for the special case of TEM-like modes with no cutoff,  the propagation constant for the channels is linear in $k$, $\beta_{pm} = n_pk$.  \autoref{Finite Element eigenvalue problem} then has the quadratic form

\begin{align} 
    \begin{pmatrix}
        \mathbf{L} - k \mathbf{C}' - k^2 \mathbf{M}  && -k\mathbf{P}'_{\mathcal{I}_{in}}\\
        \mathbf{Q}^\dagger_{\mathcal{I}_{0}} && -\mathbf{\Delta}_{\mathcal{I}_0}
    \end{pmatrix}   \begin{pmatrix}
        \mathbf{f} \\ \mathbf{a}_{\mathcal{I}_{in}}
    \end{pmatrix}  = \begin{pmatrix}
        0\\0
    \end{pmatrix},
\end{align}
where $\mathbf{P}_{\mathcal{I}_{in}} = k \mathbf{P}'_{\mathcal{I}_{in}}$ and $\mathbf{C}_{\mathcal{I}_{in}} = k \mathbf{C}'_{\mathcal{I}_{in}}$.

This quadratic eigenvalue problem can be solved by standard linearization techniques.

\section{Codimension estimate of minimal number of parameters for overconstrained problems}

\subsection{Generic systems}
In overconstrained problems, we are considering  a tall matrix $\mathcal{C} \in \mathbb{C}^{m\times n}$  with $m>n.$  For the constraint matrix to have a nontrivial solution for this case, we must have ${\rm nullity}(\mathcal{C})\geq1$. Or using the Rank-Nullity theorem, we need to have ${\rm rank} (\mathcal{C})\leq n-1$. We start by finding the codimension of the constraint matrix in terms of its rank. 
\\

\textbf{Claim}: A matrix $\mathcal{C} \in \mathbb{C}^{m\times n}$ with rank $r$ has a complex dimension $(m+n)r - r^2$.
\\

\textbf{Proof}:
Let $\mathcal{C} \in \mathbb{C}^{m\times n}$ be a matrix with rank $r$, so it must have $r$ linearly independent rows. Changing the order of the rows does not change the dimension of the matrix, so we can write $\mathcal{C} = \begin{pmatrix}
    B\\
    L
\end{pmatrix}$, where $B$ is the matrix containing the linearly independent rows of $\mathcal{C}$ and $L$ is the matrix constructed from the remaining rows. $B$ is full rank; and therefore it must have a dimension $nr$. Since every row of $L$ is a linear combination of the rows of $B$, it follows that we need $r$ complex numbers to uniquely specify each row of $L$. That is a total of $(m-r)r$ complex numbers to fully construct $L$ and to fully construct $\mathcal{C}$, we need $nr + (m-r)r = nr+ mr - r^2$ complex numbers. 
\\
\\

\textbf{Note}: The complex codimension of a matrix $\mathcal{C}\in \mathbb{C}^{m\times n}$  with rank $r$ is given by $ nm - nr - mr + r^2 = (n-r)(m-r)$. 
\\

\textbf{Note}: From the Rank-Nullity theorem, $r = n - {\rm nullity}(\mathcal{C})$ and thus the codimension of $\mathcal{C}$ is ${\rm nullity}(\mathcal{C})(m - n + {\rm nullity}(\mathcal{C}))$.
Since we want the nullity to be at least 1, we can see that in general, we need at least $(m-n+1)$ complex numbers or $2(m-n+1)$ real parameters to have a solution for the overconstrained problem at a single frequency. In our notation, this is just $2(n_D - n_N +1)$ real parameters.

\subsection{Systems with time-reversal symmetry}\label{Codimension of CPE for S-matrix with T-symmetry}

Time-reversal symmetry $(\mathcal{T})$ can reduce the codimension of perfect reflection functionalities,  which in turn reduces the minimum number of independent parameters required to realize them on the real axis. For example, the CCON $NNDD$, is $\mathcal{T}$-symmetric and has codimension 1 (rather than 2 as for generic CCONs). A similar reduction appears in overconstrained perfect reflection. This includes all cases where only $N$ and $D$ boundary conditions are considered, with $n_D > n_N$. For example, for $NDDD$, without symmetry, it would require a set of 6 parameters to be achieved at a single frequency on the real axis, but with $\mathcal{T}$, it would require only 3 parameters. We show why this is true below. 
\\

\textbf{Claim}: In systems with $\mathcal{T}$-symmetry, the number of parameters required to achieve the overconstrained functionality $(NDD\cdots D)$ at a single frequency on the real axis is $n_D$. 
\\

\textbf{Proof}: To show this we first write the scattering matrix in the form:
\begin{align}
    S =\begin{pmatrix} S_{11} & T^T\\T & R\end{pmatrix},
\end{align}
where $S_{11}$ is a scaler, $T$ is an $n_D \times 1$ column vector and $R$ is an $n_D \times n_D$ block. The constraint associated with the CCON $(NDD\cdots D)$ is $T = 0$. With time-reversal symmetry, the scattering matrix is unitary and complex symmetric, and the space of $(n_D+1) \times (n_D + 1)$ unitary and symmetric matrices has a real dimension $\frac{(n_D+1)(n_D  +2)}{2}$.  Imposing $T=0$ forces $S$ to be a block diagonal matrix:
\begin{align}
    S =\begin{pmatrix} S_{11} & 0_{1 \times n_D}\\0_{n_D \times 1} & R\end{pmatrix}.
\end{align}
Because of time-reversal symmetry, both $S_{11}$ and $R$ are unitary and complex symmetric. Thus, the dimension constrained space is given by $1 + \frac{n_D(n_D  + 1)}{2}$. The real codimesnion is given by the difference:
\begin{align}
    \frac{(n_D+1)(n_D  +2)}{2} -  \frac{n_D(n_D  + 1)}{2} -1 = n_D.
\end{align}
So the number of degrees of freedom required to achieve this functionality is $n_D$.
A special case is in the two-port $\mathcal{T}$-symmetric case of transmission zeros $(ND)$, which requires only a single parameter to be tuned to the real axis. 

We now generalize this to an arbitrary number of $N$ channels. 
\\

\textbf{Claim}: In systems with $\mathcal{T}$ symmetry, the number of parameters required to achieve the overconstrained functionality $(NN\cdots NDD\cdots D)$ at a single frequency on the real axis is $n_D - n_N + 1$. 
\\

\textbf{Proof}: With $\mathcal{T}$-symmetry, the scattering matrix $S$ is unitary and complex symmetric and could be written as:
\begin{align}
    S =\begin{pmatrix} R_1 & T^T\\T & R_2\end{pmatrix},
\end{align}
where $R_1$ is an $n_N\times n_N$ symmetric block, $T$ is an $n_D \times n_N$ matrix and $R_2$ is an $n_D \times n_D$ block.
The constraint associated with the functionality $(NN\cdots NDD\cdots D)$ is ${\rm nullity}(T) \geq 1$. 
In this space, $T$ is of rank $n_N - 1$, and so the first column could be written as a linear combination of the remaining $n_N - 1 $ columns, and using matrix operations we can reduce the constraint matrix to $T = [0_{n_D\times 1}, \tilde{T}]$, where $\tilde{T}$ is full-rank. The $n_N -1$ coefficients used in this reduction are needed to uniquely determine the constrained space. From this we can write:
\begin{align}
    S = \begin{pmatrix}R_{11} & R_{21}^T &  0_{1\times n_D}\\ R_{21} & R_{22} & \tilde{T}^T\\0_{n_D \times 1} & \tilde{T} & R_2
    \end{pmatrix}.
\end{align}
Since $S$ is unitary we have $S S^\dagger = \mathbb{I}$, which in turn gives $\tilde{T}R^*_{21} = 0.$ Because $T$ is full rank, we must have $R_{21} = 0$. From this we can write the scattering matrix as:
\begin{align}
    S = \begin{pmatrix}R_{11} & 0_{1\times (n_D + n_N-1)} &  \\ 0_{(n_D + n_N -1)} & \tilde{R}_{2} 
    \end{pmatrix}.
\end{align}
From this we can count the dimension of this constrained space. $R_{11}$ and $\tilde{R}_2$ are both unitary symmetric matrices with real dimensions $1$ and $\frac{(n_{D}+n_N - 1)(n_D+n_N)}{2}$ respectively. The dimension of the unconstrained unitary and symmetric scattering matrix is $\frac{(n_D+ n_N)(n_D + n_N + 1)}{2}$. 
So the number of required parameters for this functionality is 
\begin{align}
    \frac{(n_D+ n_N)(n_D + n_N + 1)}{2}-\frac{(n_{D} +n_N-1)(n_D + n_N)}{2} - 2(n_N - 1) = n_D - n_N +1.
\end{align}
The last term $2(n_N - 1)$ comes from the $n_N - 1 $ complex coefficients used in the reduction $T = [0_{n_D\times 1}, \tilde{T}]$.
We note that the CCON with a balanced number of $N$ and $D$ boundary conditions has codimension 1 in reciprocal lossless systems, while all other CCONs have codimension 2 under the same symmetry. In parameter space, this means that perfect reflection CCONs trace lines and other CCONs are discrete points. Fig. S2 illustrates this special case. 

\begin{figure}[!ht]\label{ }
    \centering \includegraphics[width=0.8\textwidth]{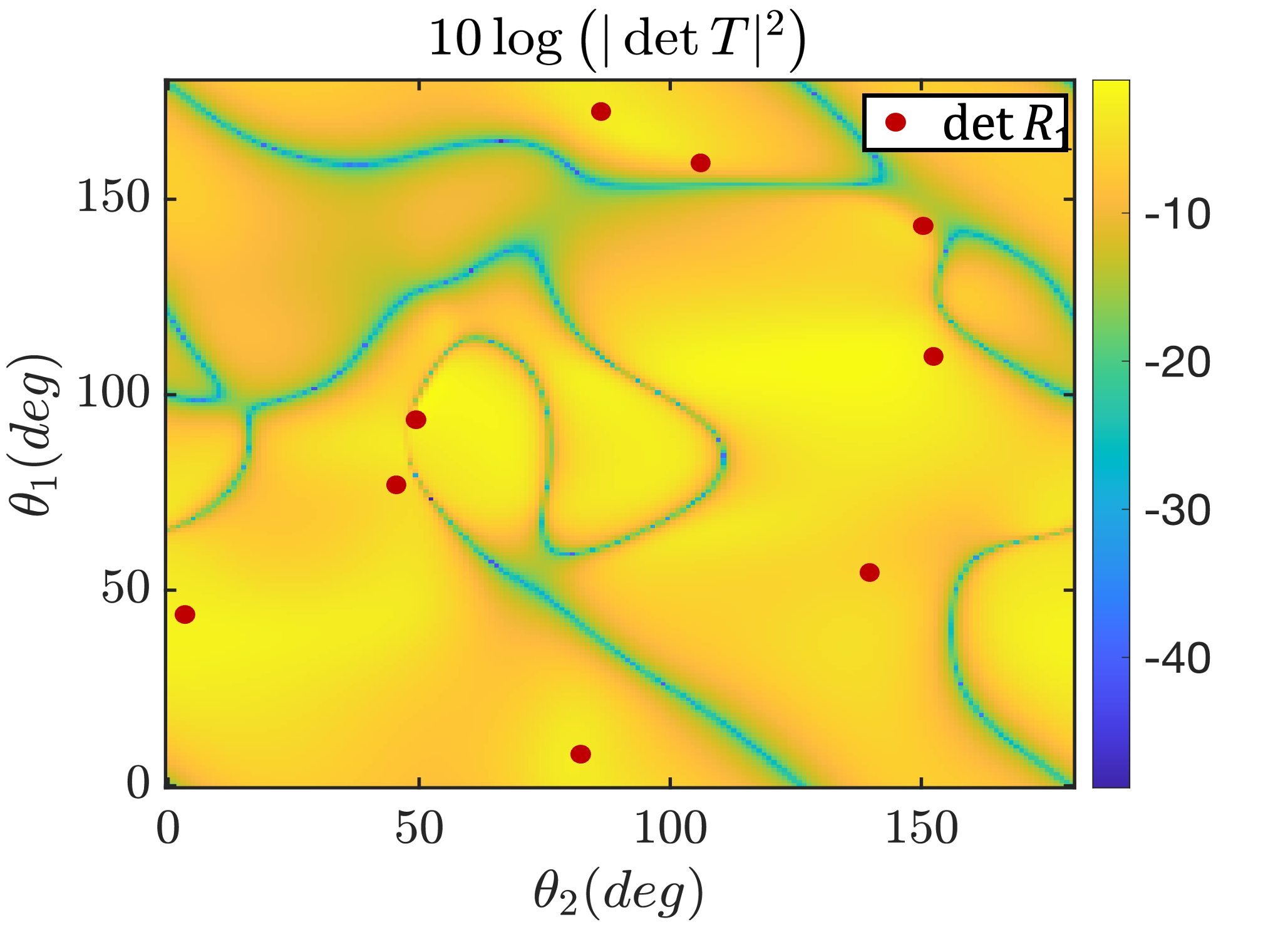}
    \caption{A heat map showing $|\det T|^2$ in ${\rm dB}$, for the same chaotic cavity in Fig. 3 in the MS. The the two dimensional parameter space of $(\theta_1, \theta_2)$ correspond to the orientation of two scatterers placed within the cavity. The red filled circles are the zeros of the two-port reflection matrix. The figure shows the distinction between the two cases, with codiminsion 1 for the two-port perfect reflection case $\det T(\omega) = 0$, and the codimension 2 reflection zeros  $\det R(\omega) = 0.$}
\end{figure}

\subsection{Case of non-chaotic cavities}
The results presented in the main text in Figs. 3, 4, 5(a-b) all used a ray-chaotic D-shaped microwave cavity, to ensure that each tunable parameter is able to change the field distribution in the entire cavity.  In this case, using elliptical scatterers of wavelength scale, we find that all tunable parameters are efficient and independent. However, it is known that scatterers inside arbitrary cavities can generate chaotic or mostly chaotic ray dynamics (an example is the Sinai billiard, a square cavity with a circular inclusion at the center, which provably has ray chaos except on sets of measure zero).  Therefore the choice of a cavity such as the D-cavity, which generates ray chaos in the absence of scatterers, is mainly to provide an extra source of chaotic scattering, to avoid any non-universal effects which might bias the system and make it less tunable. However, we were able subsequently to test whether the statistical behavior was measurably different in a rectangular cavity with similar scatterers. The results, shown in Fig. S3, show a negligible difference between the two cases, where we compare the data from the lossless rectangular cavity to the data from the D-cavity with the same number of scatterers (Fig. 5(a)).  Each cavity shows excellent agreement with codimension predictions for the minimal number of tunable parameters required to achieve various functionalities. 
Note also the results of Fig. 6, for the grating coupler model, which due to its flat surfaces should not generate ray chaos, and yet also obeys the minimal parameter predictions. These findings lead us to the conclusion that strong interchannel scattering is the key ingredient, and details of the phase space structure are less important.
\begin{figure}[h]
    \centering
    \includegraphics[width=0.95\linewidth]{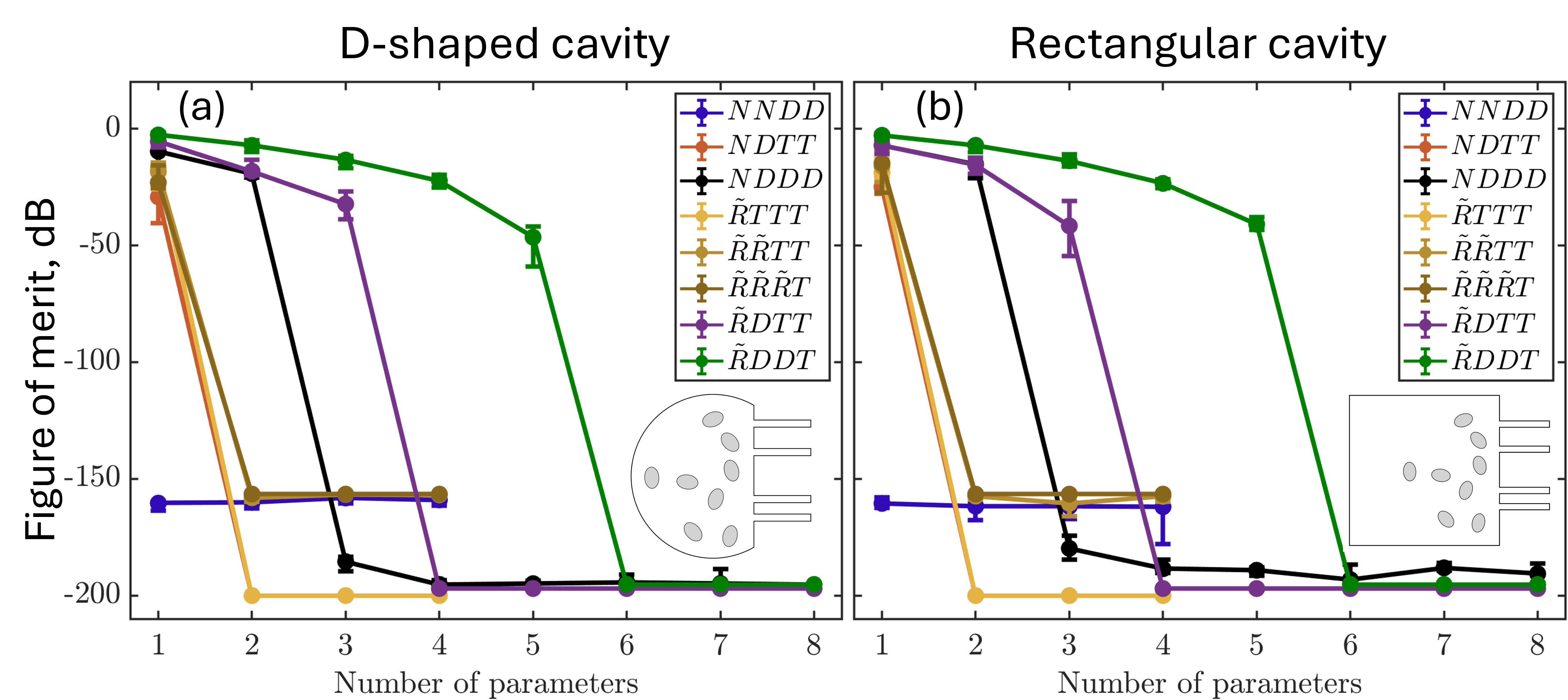}
    \caption{(a) Statistical study for the D-shaped lossless cavity, copied from Fig. 5(a) of the main text. Inset shows a generic configuration of the cavity with eight elliptical scatterers present. (b) Similar statistical study for the rectangular cavity, with inset showing a generic cavity configuration. We find that the codimension counting successfully predicts the minimum number of required tuning parameters in the case of a rectangular non-chaotic cavity, just as it did in the chaotic case.}
    \label{fig:SM_statistical_study_square_cavity}
\end{figure}

\subsection{Breakdown of minimal parameter counting in the case of abundant loss}
As we alluded to in the main text, codimension arguments prescribe necessary, but not sufficient conditions for achieving given functionalities. At a minimum, one additionally needs all the tunable parameters to be independent, but more importantly, the tunable parameters must be "strong enough": by tuning them one must be able to traverse a large enough volume in the space of S-matrices. This condition cannot be described by simple dimension counting, it relies on finer geometrical properties of the compact manifold of S-matrices realizable with a given set of parameters.

One instance where the discussion above becomes prominent is in the case of abundant loss: if we start with a cavity with an average absorption through every port of $\sim 90\%$, as we show in the figure below, the necessary conditions on the number of minimal parameters provided by the codimension argument become insufficient (while remaining necessary): one needs more parameters than the codimension predicts. For example, achieving CCONs of types $NDTT, \Tilde R\Tilde RTT,\Tilde R\Tilde R\Tilde R T$ requires 3 parameters in the case of 90\% loss instead of the 2 parameters required in the case of 50\% loss, $\Tilde RTTT$ requires 4 parameters instead of 2, $\Tilde RDTT$ requires 5 parameters instead of 4 and $\Tilde R DDT$ requires 7 parameters instead of 6. 

While we do not yet have a deeper theory capable of providing the refined sufficient conditions in this case, the theory of CCON spectra \textit{is} able to predict when this ``breakdown'' of the codimension happens. Indeed, by inspecting the complex CCON spectra for cavities with larger and larger loss, we see that adding more loss moves all the CCON spectra lower in the complex frequency plane. 
Therefore, achieving CCONs or more general CPE processes (which are intersections of CCONs) at real frequencies is harder for larger loss, as the CCONs have to move a greater distance in the complex frequency plane to appear back on the real frequency axis.

While complementing the theory with a description of the sufficient conditions in such limiting cases is desired, in scenarios of most practical importance, it is not necessary, as for example the case of \~90\% average loss through all ports would already disqualify the device as a good candidate for a reconfigurable multifunctional microwave component.

\begin{figure}[h]
    \centering
    \includegraphics[width=0.9\linewidth]{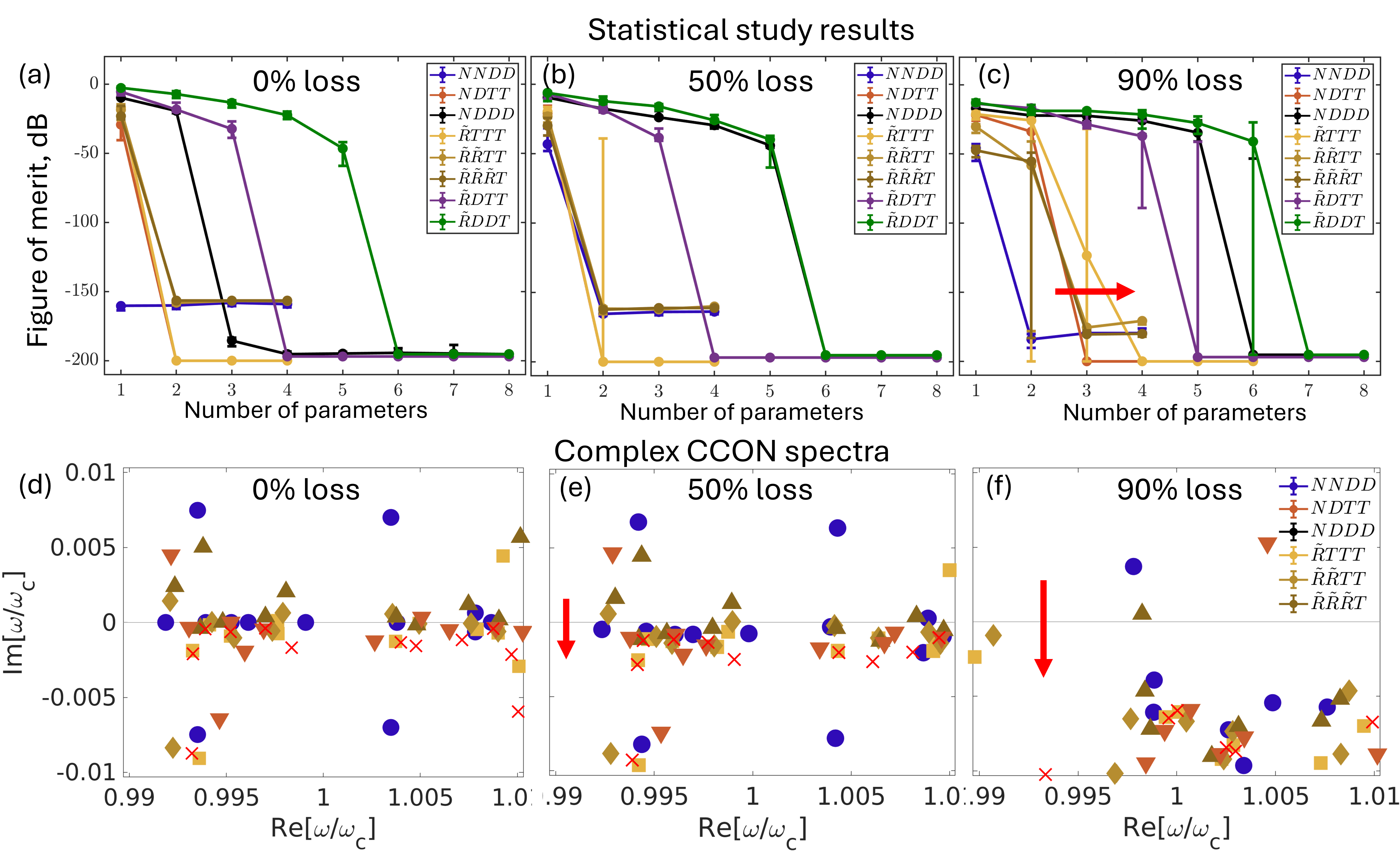}
    \caption{(a,b) The results of the statistical studies for a lossless cavity and cavity with 50\% average loss - copied verbatim from Fig. 5(a,b) of the main text. (c) The statistical study for the cavity with 90\% average loss shows that more parameters are required to achieve the same functions - the codimension is no longer a tight prescription, but is rather a lower bound on the number of parameters required. (d,e,f) The distributions of CCON complex spectra in the complex frequency plane for (d) 0\%, (e) 50\% and (f) 90\% average cavity loss, for a generic configuration of the D-shaped chaotic cavity with two elliptical scatterers. We see that with increasing loss, all the complex spectra generically move down in the complex frequency plane, making it harder to tune them back to the real axis, and thus requiring more parameters to do so. .
    }
    \label{fig:SM_CCON_figure}
\end{figure}

\section{CCONs are the building blocks of overconstrained functionalities}
When $n_D > n_N$, the constraint matrix is tall: $\mathcal{C} \in \mathbb{C}^{m\times n}$, with $m>n$. We will show that for any such matrices, if all the maximal minors vanish at a given frequency (i.e., all CCON zeros formed from the same set of incoming channels meet at a single point in the complex frequency plane), then the constraint matrix has a nontrivial nullspace at that same frequency.
\\

\textbf{Claim}: If all zeros of CCONs that can be built from the constraint matrix while sharing the same columns meet at a single point in the complex frequency plane, then the constraint matrix must have a nontrivial nullity at that same point.
\\

\textbf{Proof 1}: 
We first note that even though the constraint matrix is rectangular, we can still write the condition for existence of solutions in terms of the determinant of a single matrix, namely $\det(\mathcal{C}^\dagger \mathcal{C}) = 0$. We then use Cauchy-Binet formula on $\mathcal{C}^\dagger \mathcal{C}$ gives
\begin{align}
    \det(\mathcal{C}^\dagger \mathcal{C}) = \sum_{I \in [m], |I|=n} \det(\mathcal{C}^\dagger_{[n], I}) \det(\mathcal{C}_{I, [n]}).
\end{align}
Where $[n] = \{1, \cdots, n\}$ and $[m] = \{1, \cdots, m\}$. $\mathcal{C}_{I, [n]}$ denotes the submatrix with row indices $I$ and all the columns of $\mathcal{C}$. We can simplify the last expression to   
\begin{align}
    \det(\mathcal{C}^\dagger \mathcal{C}) = \sum_{I \in [m], |I|=n} \left|\det(\mathcal{C}_{I, [n]})\right|^2.
\end{align}
From this, it is clear to see that we have a zero of $\mathcal{C}^\dagger \mathcal{C}$ (i.e., $\mathcal{C}$ has a nontrivial nullity) if and only if all maximal minors of $\mathcal{C}$ vanish. 
\\

\textbf{Proof 2 (by contradiction)}:
Assume that all maximal minors of $\mathcal{C}$ vanish, but $\mathcal{C}$ has a trivial nullity.  Then ${\rm rank}(\mathcal{C}) = n$ which implies that there are $n$ linearly independent rows of $\mathcal{C}$. These rows form a submatrix whose determinant does not vanish, but this contradicts the assumption that all maximal minors must vanish. Thus, $\mathcal{C}$ must have a nontrivial nullity. 
\\

\textbf{Note}: For an overconstrained problem, there are $\binom{m}{n}$ CCONs that share the same incoming channels (i.e., the same columns as the the constraint matrix $\mathcal{C}$). The claim above does not specify how many such CCONs we need in general. Below, we illustrate what the claim implies in some examples.
\\

\textbf{Examples}:
\begin{align*}
    (\tilde{R}\tilde{R}DT )= (\tilde{R}\tilde{R}TT) + (N\tilde{R}DT) + (\tilde{R}NDT)
    \end{align*}
\begin{align*}
    (N\tilde{R}DD)=(N\tilde{R}DT)+(N\tilde{R}TD)+(NNDD).
\end{align*}
\begin{align*}   (\tilde{R}\tilde{R}\tilde{R}DT)=(\tilde{R}\tilde{R}\tilde{R}TT)+(\tilde{R}\tilde{R}NDT)+(\tilde{R}N\tilde{R}DT)+(N\tilde{R}\tilde{R}DT).
\end{align*}
The next claim quantifies how many  CCON coincidences suffice "in general" to show that the constraint matrix is rank-deficient. 
\\

\textbf{Claim}: Let $\mathcal{C} \in \mathbb{C}^{m\times n}$ with $m>n$. If the first $n-1$ rows of $\mathcal{C}$ are linearly independent and every $n\times n$ minor of $\mathcal{C}$ formed from all columns of $\mathcal{C}$ and the row set $\{1, \cdots n-1, i\}$, where $i\in \{n, \cdots, m\}$ vanishes then the constraint matrix must have a nontrivial nullity. 
\\

\textbf{Proof}: 
Let 
\begin{align*}
    M_k =  \begin{pmatrix}
    r_1\\\vdots\\r_{n-1}\\r_{n+k}
    \end{pmatrix}
\end{align*}
where $r_i$ is the $i_{th}$ row of 
 $\mathcal{C}$ and $k = 0, 1, \dots , m$. Assume that the first $n-1$ rows are linearly independent and that all $M_k$ have a vanishing determinant. This implies that you can write $r_{n+k}$ in terms of the first $n-1$ rows, and this is true for all $k = 0, 1, \dots, m$. It then follows that all rows $r_i$ with $i \geq n$, are in the span of the first $n-1$ columns. Thus, $\mathcal{C}$ is of rank $n-1$ and the nullity is 1. 
\\

\textbf{Note}: Because there are $m-n+1$ matrices $M_k$, and we want them all to have a vanishing determinant, it may seem that in general we need $m-n+1 = n_D - n_N +1$ CCONs to coincide to have a solution for the overconstrained problem. But it should be clear that there is no way to know in advance which $n-1$ rows are linearly independent, so the number above may not be the number of CCONs you require to coincide in general, but this does not change the codimension arguments above about the minimal number of degrees of freedom required for a given functionality. So it is possible to have a coincidence of $n_D - n_N +1$ CCONs and not have the desired functionality. This does not happen when $n_N = 1$. For $n_N = 2$, this could happen only for the trivial case where the base row is a matrix of zeros. 
\\

\textbf{Examples:}
\begin{align*}
    (\tilde{R}\tilde{R}DT) = (\tilde{R}\tilde{R}TT) + (\tilde{R}NDT)
\end{align*}
\begin{align*}
    (N\tilde{R}DD) = (N\tilde{R}DT)+(N\tilde{R}TD) 
\end{align*}

\section{R-zeros are the building blocks of perfect routing solutions}

Here we prove that for a symmetric scattering matrix, perfect routing solutions are coincidences of R-zeros in the complex frequency plane. We first define perfect routing as overconstrained functionalities containing only the boundary conditions $\Tilde{R}, T$ and $D$.
\\

\textbf{Claim}: 
Let $S$ be a symmetric matrix and let 
\begin{align*}
    \mathcal{C} = \begin{pmatrix}
        B \\ 
        d_1^T\\
        \vdots \\
        d_{m-n}^T
    \end{pmatrix}, \quad H_i = \begin{pmatrix}
        B & d_i \\
        d_i^T & c_i
    \end{pmatrix}
\end{align*}
be the reflection matrix constructed from the basis matrix  $B = [S_{st}]^{1\leq s \leq n}_{1 \leq t \leq n}$, with the first $n$ rows and columns of $S$, the $(n+i)_{th}$ column of $S$ with the first $n$ rows of $S$, $d_i = [S_{st}]^{1\leq s \leq n}_{t = n+i}$ , and $c_i = S_{n+i, n+i}$.  Then the following statements are true:
\\
\\
(i) If $B$ has nullity 1 and $H_i $ has a vanishing determinant for all $i \in  \{1, 2, \cdots, m - n\}$ then the constraint matrix $\mathcal{C}$ has nullity 1. 
\\
\\
(ii) If  $\mathcal{C}$ has a nontrivial nullity, then $B$ and $H_i$ have nontrivial nullity for all $i \in  \{1, 2, \cdots, m - n\}$ .
\\
\\

\textbf{Proof 1}:
(i) Suppose $B$ has nullity 1 and $H_i$ have a zero determinant for all  $i \in  \{1, 2, \cdots, m - n\}$, then the columns of $H_i$ are linearly dependent. In that case, either (1) $\begin{pmatrix}
    d_i \\ c_i
\end{pmatrix}$ could be written in terms of the of the other columns of $H_i$ or (2) the columns of the matrix $\begin{pmatrix}
    B \\ d_i^T
\end{pmatrix}$ are linearly dependent. If (1) then $d_i$ is in the span of the rows of $B$, and $d_i^T$ could be written in terms of the rows of $B$. If (2), then $\begin{pmatrix}
    B \\ d_i^T
\end{pmatrix}$ must be of rank at most $n-1$, but since $B$ is of rank $n-1$, it follows that $d_i^T$ is in the span of the rows of $B$. In both cases, $d_i^T$ could be written as a linear combination of the rows of $B$. This must be true for all $i \in  \{1, 2, \cdots, m - n\}$, and thus the constraint matrix $\mathcal{C}$ must be of rank $n-1$ or nullity 1. 
\\
\\
To prove (ii), suppose that $\mathcal{C}$ has nullity 1, then there exists a nonzero vector $\alpha$ such that

\begin{align*}  
\begin{pmatrix}
    B \\ d_1^T \\ \vdots \\ d_{m-n}^T
\end{pmatrix}\alpha =  \begin{pmatrix}
    B\alpha \\ d_1^T \alpha\\ \vdots \\ d_{m-n}^T \alpha
\end{pmatrix} = 0. 
\end{align*}
It is then easy to see that $\alpha$ and $\begin{pmatrix}
    \alpha \\ 0
\end{pmatrix}$
is in the Kernel of $B$ and $H_i$ respectively, and therefore both $B$ and $H_i$ must have a zero determinant. 
\\

The last claim says that for a symmetric scattering matrix $S$ with scattering boundary conditions involving only $\tilde R$ and $D$, having a nontrivial solution is generally equivalent to the coincidence of $m-n + 1 = n_D + 1$ R-zeros. 
\\

\textbf{Proof 2}: Here we prove the same result by using our notation and time-reversal symmetry arguments. We first introduce the notation $\underset{\varepsilon,\omega}{(\tilde{R}\tilde{R}\cdots \tilde{R} DD \cdots D TT \cdots  T)}$ to mean that there is a nontrivial solution satisfying the corresponding boundary conditions at $\omega$ with the scattering system described by a permittivity $\varepsilon$. We prove this in succession:
\\
\\
1. 
\color{red}
\begin{align}
    \underset{\varepsilon,\omega}{(\tilde{R}DT)} = \underset{\varepsilon,\omega}{(\tilde{R}TT)} \bigoplus \underset{\varepsilon,\omega}{(\tilde{R}\tilde{R}T)}.
\end{align}
\color{black}
\\
This simply means that $S_{11} = S_{21} = 0$ iff $S_{11} = 0$ and $\begin{pmatrix}
    S_{11} & S_{12}\\
    S_{21} & S_{22}
\end{pmatrix}
$ has a zero determinant. 
To show this, we note that   $S_{11} = S_{21} = 0$ clearly implies a zero of the two port reflection matrix $\begin{pmatrix}
    S_{11} & S_{12}\\
    S_{21} & S_{22}
\end{pmatrix}$. For the converse, we see that a zero of the two port reflection matrix implies $S_{11}S_{22} - S_{12}S_{21}= 0$. But $S_{11} = 0$ and because of symmetry we must then have $S_{21} = S_{12} = 0$. 
\\
\\
2.
\color{red}
\begin{align}     \underset{\varepsilon,\omega}{(\tilde{R}DD\cdots DT)} = \underset{\varepsilon,\omega}{(\tilde{R}TT\cdots TT)}  \underset{D \rightarrow T, D_i \rightarrow \tilde{R}}{\bigoplus} \underset{\varepsilon,\omega}{(\tilde{R}DD\cdots DT)}
\end{align}
\color{black}
The last symbol means that we take the sum over all possibilities in which all $D$'s go to $T$ except one of the $D$'s go to $\tilde{R}$. In matrix language, this says that 
  $S_{i1} = 0$ for all $i\in \{1, 2, \cdots, m\},$ iff $S_{11} = 0$ and $\begin{pmatrix}
    S_{11} & S_{1i}\\
    S_{i1} & S_{ii}
\end{pmatrix}$ has a zero determinant for all $i\in \{2, \cdots, m\}$.
\\
\\
The proof for (2) easily follows from (1) by taking the scattering elements pairwise ${S_{11}, S_{i1}}$. 
\\
\\
3. 
\color{red}
\begin{align}\label{Routing one column with one T in symbols}     \underset{\varepsilon,\omega}{(\tilde{R}DD\cdots DT)} = \underset{\varepsilon,\omega}{(\tilde{R}\tilde{R}\tilde{R}\cdots \tilde{R}T)}  \underset{D \rightarrow \tilde{R}, D_i \rightarrow T}{\bigoplus} \underset{\varepsilon,\omega}{(\tilde{R}DD\cdots DT)}.
\end{align}
\color{black}
We show this by time reversal
\color{red}
\begin{align*}
    \underset{\varepsilon,\omega}{(\tilde{R}DD\cdots DT)}& = \mathcal{T}\left[\underset{\varepsilon^*,\omega^*}{(TDD\cdots D\tilde{R})}\right]
    \\
    &= \mathcal{T}\left[\underset{\varepsilon^*,\omega^*}{(TTT\cdots T\tilde{R})} \underset{D \rightarrow T, D_i \rightarrow \tilde{R}}{\bigoplus} \underset{\varepsilon^*,\omega^*}{(TDD\cdots D\tilde{R})}\right]
    \\
    &= \underset{\varepsilon,\omega}{(\tilde{R}\tilde{R}\tilde{R}\cdots \tilde{R}T)}  \underset{D \rightarrow \tilde{R}, D_i \rightarrow T}{\bigoplus} \underset{\varepsilon,\omega}{(\tilde{R}DD\cdots DT)}
\end{align*}
\color{black}
\\
\\
4. A $T$ boundary condition does not change the constraint matrix and can always be added to equations like \autoref{Routing one column with one T in symbols}.

\color{red}
\begin{align}\label{Routing column in symbols}     \underset{\varepsilon,\omega}{(\tilde{R}DD\cdots DTT\cdots T)} = \underset{\varepsilon,\omega}{(\tilde{R}\tilde{R}\tilde{R}\cdots \tilde{R}TT \cdots T)}  \underset{D \rightarrow \tilde{R}, D_i \rightarrow T}{\bigoplus} \underset{\varepsilon,\omega}{(\tilde{R}DD\cdots DTT\cdots T)}.
\end{align}
\color{black}
\\
\\
5. \color{red}
\begin{align}
\underset{\varepsilon,\omega}{(\tilde{R}\tilde{R}\cdots \tilde{R} DD\cdots DT)}&= \underset{\varepsilon,\omega}{(\tilde{R}\tilde{R}\tilde{R}\cdots \tilde{R}TT\cdots T)}  \underset{D \rightarrow T, D_i \rightarrow \tilde{R}}{\bigoplus} \underset{\varepsilon,\omega}{(\tilde{R}\tilde{R}\cdots \tilde{R}DD\cdots DT)}.
\end{align}
\color{black}
We use time-reversal on \autoref{Routing column in symbols} 

\color{red}
\begin{align*}
    \underset{\varepsilon,\omega}{(\tilde{R}\tilde{R}\cdots \tilde{R} DD\cdots DT)}& = \mathcal{T}\left[\underset{\varepsilon^*,\omega^*}{(TT\cdots TDD\cdots D\tilde{R})}\right]
    \\
    &= \mathcal{T}\left[\underset{\varepsilon^*,\omega^*}{(TT\cdots T\tilde{R}\tilde{R}\cdots \tilde{R})} + \underset{D \rightarrow \tilde{R}, D_i \rightarrow T}{\bigoplus} \underset{\varepsilon^*,\omega^*}{(TT\cdots TDD\cdots D\tilde{R})}\right]
    \\
    &= \underset{\varepsilon,\omega}{(\tilde{R}\tilde{R}\tilde{R}\cdots \tilde{R}TT\cdots T)} + \underset{D \rightarrow T, D_i \rightarrow \tilde{R}}{\bigoplus} \underset{\varepsilon,\omega}{(\tilde{R}\tilde{R}\cdots \tilde{R}DD\cdots DT)}
\end{align*}
\color{black}
\\
\\
6. The last thing we do is that we add a general number of $T$  channels  to the end which does not change the constraint matrix
\color{red}
\begin{align}
\underset{\varepsilon,\omega}{(\tilde{R}\tilde{R}\cdots \tilde{R} DD\cdots DTT\cdots T)}&= \underset{\varepsilon,\omega}{(\tilde{R}\tilde{R}\tilde{R}\cdots \tilde{R}TT\cdots TT\cdots T)}  \underset{D \rightarrow T, D_i \rightarrow \tilde{R}}{\bigoplus} \underset{\varepsilon,\omega}{(\tilde{R}\tilde{R}\cdots \tilde{R}DD\cdots DTT\cdots T)}.
\end{align}
\color{black}
This proves the statement.

\section{Quantum Graphs}
In Quantum Graphs~\cite{Kottos1997QuantumGraphs, Kottos2003QuantumScattering}, the system consists of a set of vertices $V$ connected by edges $E$. Along each edge, the field satisfies the 1D Helmholtz equation. At each vertex, boundary condition are imposed such that the field is continuous (Neumann condition) and the sum of the outgoing derivatives vanishes (Kirchhoff condition). Despite their simplicity, these system were shown to have some statistical properties that are characteristic of  chaotic cavities \cite{Kottos1997QuantumGraphs}. From the governing equations, a compact expression for the scattering matrix can be derived:

\begin{align}
    S(k) = \mathbb{I} - 2i W^T \frac{1}{H(k) + iW W^T}W ,  
\end{align}
where 
\begin{align}
    H_{\alpha \beta} = \begin{cases} - \sum_{\mu\neq \alpha} C_{\alpha \mu} \cot(kL_{\alpha\mu} + \phi_{\alpha \mu}) & \text{if} \quad \alpha = \beta\\
    C_{\alpha\beta} e^{-iA_{\alpha\beta}L_{\alpha\beta}}\csc(kL_{\alpha\mu} + \phi_{\alpha \beta}) & \text{if} \quad \alpha \neq \beta
    \end{cases}.
\end{align}
$W_{ij}$ is the coupling between the leads and vertices and is taken to be 1 when they are connected and $0$ otherwise. $C_{\alpha\beta}$ is $1$ when the vertices $\alpha$ and $\beta$ are connected, and $0$ otherwise and $A_{\alpha\beta}$ is the magnetic potential on the edge connecting the vertices $\alpha$ and $\beta$. 
\\
We test the minimal number of degrees of freedom predicted by the theory for these functionalities using a Quantum Graphs model with 4 leads and 10 connected vertices. For each functionality, we define the cost function as the smallest singular value of the constraint matrix. 

\begin{algorithm}[H]
\caption{Testing the Minimal Number of Degrees of Freedom Using Quantum Graphs}

\KwIn{Maximum number of degrees of freedom $D_{\max}$,
number of generated ensembles $N_{\mathrm{ens}}=50$,
number of initialization points for local minima
$N_{\mathrm{int}}=50$, and the connectivity matrix $C$}

\KwOut{Median figure of merit (FOM) over the ensembles}

\For{$d \leq D_{\max}$}{

    \While{$i \leq N_{\mathrm{ens}}$}{

        \tcc{Initialize}
        Choose a random wavenumber $k$\;
        Choose $N_{\mathrm{edges}}$ random bond lengths\;
        Choose $d$ random phases along $d$ randomly selected bonds\;
        Assign random phases to the remaining
        $N_{\mathrm{edges}}-d$ bonds\;

        \BlankLine
        \tcc{Local search using a quasi-Newton algorithm}

        \While{$j \leq N_{\mathrm{int}}$}{
            Use a quasi-Newton method to find a local minimum\;
        }

        Select the best (smallest) minimum over all
        initializations $j$\;

        \BlankLine
        \tcc{Post-processing}

        Compute the median FOM over the ensembles\;

    } 

} 

\end{algorithm}

\begin{figure}[!ht]\label{ fig}
    \centering \includegraphics[width=0.8\textwidth]{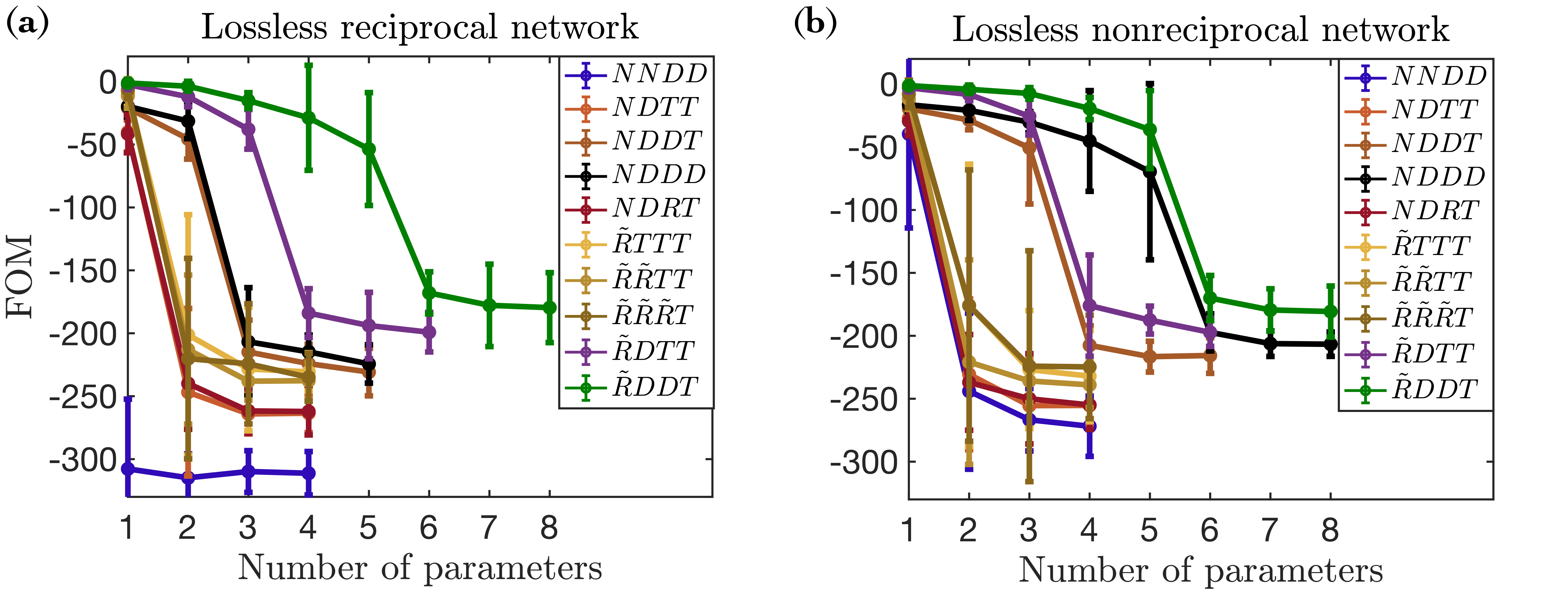}
    \caption{Minimal parameter counts from quantum graphs tests. Using a graph with 10 fully connected vertices, and 4 leads, we verify the predicted DOF for the selected functionalities for both a lossless reciprocal network (a) and a lossless nonreciprocal network (b).   1 degree of freedom for the $\mathcal{T}$-symmetric CCON $(NNDD)$ in the reciprocal case (a) and 2 in the nonreciprocal case (b) and 2 degrees of freedom all other CCONs in both cases. Overconstrained $\mathcal{T}$-assisted functionalities: 3 degrees of freedom for $(NDDT)$ and $(NDDD)$ for the reciprocal network while the nonreciprocal network requires 4 and 6 degrees of freedom respectively for the same functionalities (see~\autoref{Codimension of CPE for S-matrix with T-symmetry}). $(\tilde{R}DTT)$ or partial routing is a 4 parameter functionality and $(\tilde{R}DDDT)$ is full routing which is a 6 parameter functionality for both cases.}
\end{figure}

\section{Topological properties of CCONs}
Generically, CCONs are zeros of a complex valued function (i.e., $\det (\mathcal{C}(\omega))$), and depending on the context, they can be viewed as zeros in a general parameter space, or in the complex frequency plane. Here, we clarify the distinction between the two representations.  We can write the scattering matrix with its explicit dependence on both frequency and some general parameters $p$ that depend on the scattering system as $S(\omega, p)$, where $p$ is a $d$ dimensional vector. CCONs can then be explored in two ways: fixing $p$ and observing the discrete spectra in the complex plane or by fixing the frequency and finding the zeros in a chosen 2 dimensional parameter space. The former was the view used in the main paper, while the latter was considered in \cite{Guo2023SingularMatrices}. 
\\
\\
Without symmetry, and assuming smooth dependence on $\omega$ and  $p$, the theory of Sard and Brown~\cite{Milnor1965TopologyViewpoint, Roe2015WindingAnalysis} guarantees that those zeros are isolated points. 
\\
For now we consider those zeros in a 2-dimensional space, either a complex plane, or a parameter plane $(p_1, p_2)$. In this space we can define the function $f(s) = \det(\mathcal{C}(s))$ as a complex valued function of $s$. Around a simple zero $s_0$, the winding number is 
\begin{align}
    \text{wind}(f; s_0) = \frac{1}{2 \pi i}\int_0^{2\pi} \frac{f'(\phi)}{f(\phi)}d\phi
.\end{align}
To evaluate this, we write $f(s)$ as a two component vector with $f(s) = \left(\text{Re}\{f(s)\}, \text{Im}\{f(s)\}\right)^T$ and linearize around $s_0$: $f(s) = Df(s_0)(s - s_0)$, where $Df(s_0)$ is the Jacobian matrix at $s_0$ and is given by 
\begin{align}
    Df = \begin{pmatrix}
        \frac{\partial f_1}{\partial x}  & \frac{\partial f_1}{\partial y} \\
        \frac{\partial f_2}{\partial x}  & \frac{\partial f_2}{\partial y}
    \end{pmatrix}_{s_0}.
\end{align}
After some calculations, we arrive at
\begin{align}
    \text{wind}(f; s_0) = \text{sign}(\det (Df(s_0))).
\end{align}
So every simple zero has a winding number of $\pm 1$. 
When $f$ is analytic at $s_0$, the Cauchy-Riemann equations force  $\det(Df)\geq 0$, which ensures that
\begin{align}
    \text{wind}(f; s_0) = 1, \qquad \text{for $f$ analytic at $s_0$}.
\end{align}
Hence, within a region of the complex frequency plane where $S$ is analytic, the winding number associated with a simple zero is always $+1$. This property extends to other physical quantities for which $S$ is meromorphic (e.g., $\varepsilon, \mu$). Consequently, CCONs cannot annihilate pairwise in the complex frequency plane under continuous parametric variations of the geometry and are therefore conserved in that sense. However, annihilation between a  pole and a zero remains possible. 
\begin{figure}[H]
    \centering   \includegraphics[width=0.8\textwidth]{Figure_6_SM_v2.png}
    \caption{Visualization of CCONs in a geometric space derived from a quantum graphs calculation. (a) Phase of $(S_{21})$ showing that around the singular points the phase winds once either clockwise or counter-clockwise, corresponding to a winding number of -1 or +1. (b) The system is slightly perturbed from (a) and we observe the flow of zeros until an annihilation event occurs at the marked location. (c) Flow of zeros in geometric space  as  $kL_{13}$ is tuned from $2.8683$ (panel a) to $2.9256$ (panel b). The red star indicates the annihilation point of the zeros. Inset: schematic of the model used. }
\end{figure}
An additional property of CCONs arises when the scattering matrix is periodic in 2 parameters. By tracing a simple contour around a single unit cell, the winding number must vanish. Thus, CCONs in that space must always come in pairs with opposite winding numbers. Since our control parameters are the angles of ellipsoidal scatterers with respect to a fixed axis, they are indeed periodic and the relevant parameter spaces exhibit this property. All these properties have a natural extension to the overconstrained functionalities, where the space in which zeros exist is higher dimensional.  A complete treatment of the topological properties including such cases will appear in a future publication.



\section{Optimization}
In all of the designs, we aim to have a nontrivial nullspace for the constraint matrix. We consider the case of a square and nonsquare constraint matrix. We state the cost function used in all of the implementations.

 \subsection{Cost function}\
 In this case, $n_D = n_N$ and we need to have a zero determinant for the constraint matrix $\mathcal{C}$
 \begin{align}
     CF = |\det(\mathcal{C})|^2
 \end{align}
We can also formulate this in terms  the singular values. 

 \subsection{Tall constraint matrix}\label{Tall constraint matrix}
In this case $\mathcal{C} 
\in \mathbb{C}^{m\times n}$ with $m > n$, is non-square and we can not define the determinant directly. But we note that $N(\mathcal{C}) = N(\mathcal{C}^\dagger \mathcal{C})$,  where $N()$ refers to the nullspace, and since $\mathcal{C}^\dagger \mathcal{C}$ is square, it follows that we can formulate the optimization problem in terms of the determinant of $\mathcal{C}^\dagger \mathcal{C}$
\begin{align}
    CF = |\det(\mathcal{C}^\dagger \mathcal{C})|^2.
\end{align}
\\
Another approach is to use the singular values of the constraint matrix. If the set ${\sigma_n}$ of singular values of $\mathcal{C}$, is arranged in increasing order, then one simple cost function could be
\begin{align}
    CF = \sigma_1^2.
\end{align}

\subsection{Testing the minimal number of degrees of freedom}
The method we use in all of the optimizations is a combination of Newton's method and Linear Simplex method.

To study the dependence the optimization on the number of scatterers, we used the following procedure: \\
The scattering matrix will depend on the location of the scatterers $x_i$ inside the cavity, their orientation $\theta_i$ and driving frequency $\omega$. More explicitly we can write the scattering matrix as $ S(x_i, \theta_i, \omega)$. We aim to study the dependence of our figure of merits on the number of degrees of freedom, which would be the orientation $\theta_i$ of the scatterers. 
\begin{algorithm}[H]
\caption{Testing the Minimal Number of Degrees of Freedom in a Chaotic Cavity}

\KwIn{Maximum number of degrees of freedom $D_{\max}$,
number of generated ensembles $N_{\mathrm{ens}}$,
number of initialization points $N_{\mathrm{int}}$,
and cutoff frequency $f_c$}

\KwOut{Median figure of merit (FOM) over the ensemble}

\For{$d \leq D_{\max}$}{

    \While{$i \leq N_{\mathrm{ens}}$}{

        \tcc{Initialize}
        Choose $d$ arbitrary locations for the scatterers\;
        Choose a random frequency
        $f \in [0.96f_c,0.98f_c]$\;
        Relabel the ports\;

        \BlankLine
        \tcc{Local search using a quasi-Newton algorithm}

        \While{$j \leq N_{\mathrm{int}}=5d$}{
            Use the quasi-Newton method to find a local minimum\;
        } 

        Select the best (smallest) minimum over all
        initializations $j$\;

        \BlankLine
        \tcc{Refine}

        Use the Nelder--Mead simplex method, initialized at the
        best minimum found over $j$\;

    } 

    \BlankLine
    \tcc{Post-processing}

    Compute the median FOM over the ensemble\;

} 

\end{algorithm}

\section{Full wave simulations}
The 2D chaotic cavity used in the main text is a D-shaped structure with $R = 5.75 \lambda_c$, where $\lambda_c = c_0/f_c$ denotes the wavelength corresponding to the cutoff frequency and $c_0$ is the speed of light in vacuum. The cavity is truncated from a circular disk by a straight cut at a distance $d = R/2$ from the center and is connected to a few single mode waveguides along this cut. Inside the cavity, the scatterers consist of elliptical disks with major and minor semi-axes of $0.75\lambda_c$ $0.5 \lambda_c$, respectively. All the boundaries are perfectly conducting. Simulations in Figs. 1-4 were performed using the commercial finite-element method software COMSOL Multiphysics. Full wave simulations in Fig. 5 were performed using the boundary integral equations (BIE) package chunkIE, see~\cite{chunkie}. Full wave simulations of the periodic structures in Fig. 6 were performed using the rigorous coupled wave analysis (RCWA) package RETICOLO, see~\cite{reticolo}. Example codes used for the analyses presented in this work are available at~\cite{Alhulaymi_Coherent_Control_of_2026}.

\bibliography{references-2}


\end{document}